\documentclass[preprint,3p,sort&compress]{elsarticle}
\usepackage{amsmath,bm}
\usepackage{physics}
\usepackage{nomencl}
\usepackage{framed} 
\usepackage{amsfonts}
\usepackage{amssymb}
\usepackage{siunitx}
\usepackage{tikz}
\usepackage[version=3]{mhchem}
\usepackage{mathtools}
\usepackage{lineno}
\usepackage{setspace}
\usepackage[colorlinks,citecolor=blue,urlcolor=blue,linkcolor=blue]{hyperref}
\journal{Progress in Energy and Combustion Science}
\begin{document}
\begin{frontmatter}
\title{Lattice Boltzmann methods for combustion applications}
\author[inst1]{Seyed Ali Hosseini}
\ead{shosseini@ethz.ch}
\affiliation[inst1]{organization={Department of Mechanical and Process Engineering, ETH Zurich},
            city={Zurich},
            postcode={8092}, 
            country={Switzerland}}
\author[inst2]{Pierre Boivin}
\affiliation[inst2]{organization={Aix Marseille Universite, {CNRS}, Centrale Marseille, {M2P2}},
            city={Marseille},
            postcode={13451}, 
            country={France}}
\author[inst3]{Dominique Th\'evenin}
\affiliation[inst3]{organization={Laboratory of Fluid Dynamics and Technical Flows, University of Magdeburg ``Otto von Guericke''},
            city={Magdeburg},
            postcode={D-39106}, 
            country={Germany}}
\author[inst1]{Ilya Karlin}
\begin{abstract}
    The lattice Boltzmann method, after close to thirty years of presence in computational fluid dynamics has turned into a versatile, efficient and quite popular numerical tool for fluid flow simulations. The lattice Boltzmann method owes its popularity in the past decade to its efficiency, low numerical dissipation and simplicity of its algorithm. Progress in recent years has opened the door for yet another very challenging area of application: Combustion simulations. Combustion is known to be a challenge for numerical tools due to, among many others, the large number of variables and scales both in time and space, leading to a stiff multi-scale problem. In the present work we present a comprehensive overview of models and strategies developed in the past years to model combustion with the lattice Boltzmann method and discuss some of the most recent applications, remaining challenges and prospects.
\end{abstract}
\end{frontmatter}

\onehalfspacing
\tableofcontents\vspace{\baselineskip} 
\section{Introduction}
The lattice Boltzmann (LB) method, proposed in the early 80's has grown popular over the past decades~\cite{higuera_lattice_1989,higuera_boltzmann_1989}. The rapid emergence of this numerical method is mainly due to the simplicity and strict locality of the involved time-evolution operators~\cite{guo_lattice_2013,kruger_lattice_2017}. The locality of the operators and intrinsic coupling between the pressure and velocity fields through the distribution function (as opposed to pressure-based incompressible or low Mach solvers) allows for better performances on parallel clusters and a much more efficient treatment of flows in complex geometries \cite{kruger_lattice_2017}.
During the past decade, the LB method originally proposed for computational fluid dynamics (CFD) has been extended to many complex flow configurations ranging from non-Newtonian \cite{gabbanelli_lattice_2005,boyd_second-order_2006,ouared_lattice_2005,wang_lattice_2009,hosseini_lattice_2022}, to multi-phase \cite{shan_multicomponent_1995,mazloomi_entropic_2015,hosseini_towards_2022,fakhari2010phase,fakhari_mass-conserving_2016,fakhari_improved_2017,hosseini_lattice_2021}, and multi-component flows. Although initially limited to low-Mach isothermal flows with an ideal gas equation of state, the LB approach was later modified to lift many of these restrictions. Releasing the restriction on thermo-compressibility is an essential step to develop LB solvers for many applications such as combustion.\\
The topic of combustion modeling with LB was first touched upon in 1997 in an article by Succi et al. \cite{succi1997lattice}. Since then, and up until very recently, a limited number of publications had appeared on the topic, all limited to simplified 1-D and 2-D test-cases, see for instance~\cite{yamamoto_simulation_2002,yamamoto_combustion_2005,chen_novel_2007,chen_simple_2008,filippova_novel_2000,chiavazzo_combustion_2009,chiavazzo_coupling_2010}. The limited progress of the lattice Boltzmann method during that period might be attributed to a number of factors such as the absence of a good compressible realization, persistent issues with stability of solvers, and the absence of multi-species formulations. During the past years a considerable amount of research work has been conducted to extend the lattice Boltzmann method to compressible flows, which has led to a number of stable and efficient realizations, see for instance~\cite{feng_compressible_2016,frapolli_entropic_2015,saadat_extended_2021,dorschner_particles_2018}. In parallel, the stability domain of lattice Boltzmann solvers both for incompressible and compressible flows has been considerably expanded through more advanced collision models, see for instance~\cite{lallemand_lattice_2003,karlin_gibbs_2014,geier_cascaded_2006,geier_cumulant_2015,malaspinas_increasing_2015,hosseini_development_2020}. These two factors along with the development of models for species transport and the idea of hybrid solvers taking advantage of classical numerical methods for the species end energy balance equations led to considerable progress in combustion simulation with the lattice Boltzmann method in recent years. Contrary to the first wave of models, the more recent efforts have been extended and used for many complex configurations involving thermo-acoustics, complex geometries and turbulent flows. It has to be noted that in parallel with efforts to develop lattice Boltzmann-based models for combustion simulations, a number of attempts at developing discrete velocity Boltzmann-based models with Eulerian discretization in physical space have also been reported, see for instance~\cite{lin_double-distribution-function_2016,lin_multi-component_2017,lin_discrete_2017,lin_discrete_2019}.\\
In the present contribution we will review developments in the area of lattice Boltzmann simulations of combustion. Different challenges, solutions and models developed in that area in the past years will be presented and discussed. The review starts with a brief overview of basic concepts, i.e. target macroscopic system and basic concepts from the lattice Boltzmann method. In the third section of this review we will discuss topics specific to combustion simulations, i.e. strategies to solve the energy balance equation, models developed for species transport equations, and introduction of compressibility effects into the lattice Boltzmann solver. The review closes with section four where key points are briefly listed and future prospects and challenges are discussed.
\section{Basic concepts}
\subsection{Brief overview of target macroscopic system}
Throughout the manuscript, the target set of macroscopic equations is the multi-component system of Navier-Stokes-Fourier equations (see, e.g. \cite{poinsot_theoretical_2005})
\begin{align}
    \pdv{\rho}{t} + \pdv{\rho u_\beta}{x_\beta} &= 0, \label{eq:mass}\\ 
    \pdv{\rho u_\alpha}{t} + \pdv{\rho u_\alpha u_\beta + p \delta_{\alpha \beta}}{x_\beta} &= \pdv{\tau_{\alpha \beta}}{x_\beta}, \label{eq:momentum}\\
\pdv{\rho  E}{t} + \pdv{\rho u_\beta (E+p/\rho)}{x_\beta} &=  \pdv{\tau_{\alpha \beta} u_\alpha}{x_\beta} - \pdv{q_\beta}{x_\beta}, \label{eq:energy}\\ 
    \pdv{\rho Y_k}{t} + \pdv{\rho  u_\beta Y_k}{x_\beta} &= \pdv{ \rho V_{k, \beta} Y_k}{x_\beta}+ \dot{\omega}_k. \label{eq:species}
    \end{align}
Here $u_{\alpha}$ is the $\alpha^{\text{th}}$ component of the fluid velocity, $\rho$ is the mixture density, $E$ is the total energy (sum of internal energy $e$ and kinetic energy $u_\alpha^2/2$),  $Y_k$ is the mass fraction of species $k$, and $\delta_{\alpha\beta}$ is the Kronecker symbol (1 if $\alpha=\beta$, 0 else). 
The above system is fully closed upon choosing
\begin{description}
    \item[Equation of state:] a thermodynamic closure, linking state variables $p$, $\rho$, $e$, $T$ and $Y_k$, e.g. following the perfect gas assumption $p=\rho.\bar{r}.T=\rho \frac{\mathcal{R}}{\bar{W}}T$.
    \item[Transport models:] to define the species diffusion velocities $V_{k,\beta}$,  heat flux term $q_\beta$ and viscous stress tensor $\tau_{\alpha\beta}$.
    \item[Chemistry model:] to define the reaction rates $\dot \omega_k$.
\end{description}
\subsection{Isothermal lattice Boltzmann for incompressible flows}
The construction of a discrete kinetic solver like the lattice Boltzmann method has two main ingredients: (a) Reduction of the particles' speed continuous space to a discrete set, and (b) discretization of the resulting system of hyperbolic equations in physical space and time. In this section these two components will be briefly reviewed.
\subsubsection{Discrete velocity system and discrete equilibrium state}
The rationale behind the construction of the lattice Boltzmann method consists in using a truncated version of the Boltzmann equation with a linear approximation to the collision term to recover the dynamics of the macroscopic equations of interest, here the isothermal Navier-Stokes and continuity equations.

\paragraph{From Boltzmann-BGK to the discrete velocity Boltzmann equations}
Consistent with the terminology of the early literature, in the context of the present work we will refer to all methods using a form of the Boltzmann equation with a discrete set of particles' velocities as discrete-velocity models (DVM). In recent years interest in such models has been revived in the form of numerical methods such as the lattice Boltzmann equation. DVM generally aim at approximating the distribution function with quadrature rules or similar integral approximations and using a discrete set of velocities:
\begin{equation}
	\mathcal{V}:=\{\bm{c}_i\in \mathbb{R}^D \},
\end{equation}
changing the Boltzmann-Bhatnagar-Gross-Krook (BGK) equation~\cite{bhatnagar_model_1954} into a set of coupled hyperbolic partial differential equations:
\begin{equation}
	\frac{\partial f_i}{\partial t} + c_{i\alpha}\frac{\partial f_i}{\partial x_\alpha} = \frac{1}{\tau}\left(f_i^{\rm eq} - f_i\right).
\end{equation}
Constraints on the discrete equilibrium function, $f^{\rm eq}_i$, e.g. moments of the equilibrium distribution function to be correctly recovered, are identified via a Chapman-Enskog (CE) multi-scale expansion. For the continuity equation:
\begin{equation}
	\frac{\partial\Pi_0^{\rm eq}}{\partial t } + \frac{\partial \Pi_\alpha^{\rm eq}}{\partial x_\alpha} = 0, 
\end{equation}
while for the momentum balance equations:
\begin{equation}
	\frac{\partial\Pi_\alpha^{\rm eq}}{\partial t } + \frac{\partial \Pi_{\alpha\beta}^{\rm eq}}{\partial x_\beta} - \frac{\partial}{\partial x_\beta}\tau\left[ \frac{\partial \Pi_{\alpha\beta}^{\rm eq}}{\partial t} + \frac{\partial \Pi_{\alpha\beta\gamma}^{\rm eq}}{\partial x_\gamma}\right] = 0,
\end{equation}
where we have made use of the following notation:
\begin{equation}
	\Pi_{\alpha_1,\dots,\alpha_n} = \int \prod_{\alpha=\alpha_1}^{\alpha_n} v_\alpha  f(\bm{v},\bm{x},t) d\bm{v},
\end{equation}
meaning that for the system of interest one needs to correctly recover moments of orders zero through three of the equilibrium distribution function.\\
In the specific context of the lattice Boltzmann method, the Gauss-Hermite quadrature is utilized to satisfy most of the above-listed conditions on the discrete distribution functions, i.e.
\begin{equation}
	\int P^M\left(\bm{v}, \rho, \bm{u}\right) w\left(\bm{v}\right)d\bm{v}\cong \sum_{i=0}^{Q-1} w_i P^M\left(\bm{c}_i, \rho, \bm{u}\right),
\end{equation}
where $P^M\left(\bm{v}, \rho, \bm{u}\right)$ is a polynomial of order $M$ of $\bm{v}$ and $w(\bm{v})$ is a function of the form:
	\begin{equation}
	w\left(\bm{v}\right) = {\left(2\pi\right)}^{-D/2} \exp\left({-\frac{ {\bm{v}}^2}{2}}\right).
\end{equation}
For the quadrature to be applicable the distribution function must be expanded as the product of a polynomial series and $w(\bm{v})$ and the integration variable, i.e. $\bm{v}$ normalized via the reference temperature, i.e. $\bar{r}T_0$. A change of variable as $\bm{v}'=(\bm{v}-\bm{u})/\sqrt{\bar{r}T}$ like Grad's expansion is not possible here as it would lead to changing discrete particle velocities that are also not necessarily space-filling.\\
Choosing the abscissae, i.e. $\bm{c}_i$ to be the roots of the Hermite polynomial of order $Q$ and the weights as:
\begin{equation}
    w_i = \frac{Q!}{{\mathcal{H}_{Q-1}(\bm{c}_i)}^2},
\end{equation}
results in the maximum algebraic degree of precision, i.e. $2Q-1$. This means that the quadrature guarantees exact recovery of moments up to order $\frac{2Q-1}{2}$.\\
In the case of the classical lattice Boltzmann stencil, a third-order quadrature is used, i.e. $Q=3$ in 1-D, with:
\begin{equation}
    c_i \in \{-\sqrt{3\bar{r}_0 T_0}, 0, \sqrt{3\bar{r}_0 T_0}\},
\end{equation}
and
\begin{equation}
    w_i \in\{\frac{1}{6}, \frac{2}{3}, \frac{1}{6}\}.
\end{equation}
The simplest multi-dimensional extension of this quadrature can be obtained by operating a tensorial product of the 1-D lattice.
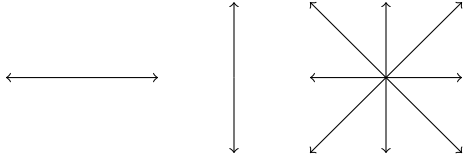
\begin{figure}
\centering
\begin{tikzpicture}
  \draw[->] (0,0) -- (0,1); 
  \draw[->] (0,0) -- (0,-1); 
  \draw[->] (-2,0) -- (-1,0);
  \draw[->] (-2,0) -- (-3,0);
    \draw[->] (2,0) -- (2,1);
    \draw[->] (2,0) -- (2,-1);
    \draw[->] (2,0) -- (1,0);
    \draw[->] (2,0) -- (3,0);
    \draw[->] (2,0) -- (3,1);
    \draw[->] (2,0) -- (1,1);
    \draw[->] (2,0) -- (1,-1);
    \draw[->] (2,0) -- (3,-1);
\end{tikzpicture}
\caption{Illustration of the tensorial product process to build D2Q9 lattice from D1Q3.}
\label{fig:tens_prod_latt}
\end{figure}
For instance, in 2-D as illustrated in Fig.~\ref{fig:tens_prod_latt}, this leads to the D2Q9 lattice with:
\begin{equation}
    c_{ix}/\sqrt{3\bar{r}T_0} \in \{0,1,0,-1,0,1,-1,-1,1\},
\end{equation}
and
\begin{equation}
    c_{iy}/\sqrt{3\bar{r}T_0} \in \{0,0,1,0,-1,1,1,-1,-1\},
\end{equation}
and
\begin{equation}
    w_i\in\{\frac{4}{9},\frac{1}{9},\frac{1}{9},\frac{1}{9},\frac{1}{9},\frac{1}{36},\frac{1}{36},\frac{1}{36},\frac{1}{36}\}.
\end{equation}
A similar procedure is used to obtain the D3Q27 lattice for 3-D simulations.
\paragraph{Discrete equilibrium: polynomial form}
A number of different ways for constructing the discrete equilibrium have been proposed over the years. One of the early approaches, first discussed in \cite{he_theory_1997} was to re-write the equilibrium as:
\begin{equation}
    f^{\rm eq} = \rho\left({2\pi \bar{r}_0 T_0}\right)^{-D/2} \exp\left\{\frac{-\bm{v}^2}{2\bar{r}_0 T_0}\right\} \exp\left\{\frac{-\bm{u}^2+\bm{u}\cdot\bm{v}}{2\bar{r}_0 T_0}\right\},
\end{equation}
and Taylor-expand the last term around Mach Ma$=0$, i.e.
\begin{equation}
    \exp\left\{\frac{-\bm{u}^2+\bm{u}\cdot\bm{v}}{2\bar{r}_0 T_0}\right\} = 1 + \frac{\bm{v}\cdot\bm{u}}{\bar{r}_0 T_0} + \frac{{(\bm{v}\cdot\bm{u})}^2}{2\bar{r}_0^2T_0^2} - \frac{\bm{u}^2}{2\bar{r}_0T_0} + \mathcal{O}\left({\|\bm{u}\|}^3/{\bar{r}_0T_0}^{3/2}\right),
\end{equation}
ultimately leading to, after discretization of particles velocity space and application of the Gauss-Hermite quadrature, the second-order polynomial discrete equilibrium:
\begin{equation}
    f_i^{\rm eq} = w_i \rho \left(1 + \frac{\bm{c}_i\cdot\bm{u}}{c_s^2} + \frac{{(\bm{c}_i\cdot\bm{u})}^2}{2c_s^4} - \frac{\bm{u^2}}{2 c_s^2}\right).
\end{equation}
An alternative construction based on an expansion of the distribution function with Hermite polynomial was proposed, which led to the following final form:
\begin{equation}
    f_i^{\rm eq} = w_i \rho \sum_{n=0}^{N} \frac{1}{n!{c_s}^{2n}} \mathcal{H}_n(\bm{c}_i):a^{\rm eq}_n(\rho,\bm{u}),
\end{equation}
where $\mathcal{H}_n$ and $a^{\rm eq}_n$ are tensors of rank $n$ representing respectively the order $n$ Hermite polynomial and coefficient.\\
Alternatively the polynomial equilibrium can also be constructed via the product form. The product form of the equilibrium distribution function (EDF) is a special realization of the moments matching approach. Considering the standard discrete velocity set D3Q27, where D=3 stands for three dimensions and Q=27 is the number of discrete velocities,
    \begin{equation}\label{eq:d3q27vel}
    	\bm{c}_i=(c_{ix},c_{iy},c_{iz}),\ c_{i\alpha}\in\{-1,0,1\},
    \end{equation}
one first defines a triplet of functions in two variables, $\xi_{\alpha}$ and $\zeta_{\alpha\alpha}$, 
    \begin{align}
    	&	\Psi_{0}(\xi_{\alpha},\zeta_{\alpha\alpha}) = 1 - \zeta_{\alpha\alpha}, 
    	\label{eqn:phi0}
    	\\
    	&	\Psi_{1}(\xi_{\alpha},\zeta_{\alpha\alpha}) = \frac{\xi_{\alpha} + \zeta_{\alpha\alpha}}{2},
    	\label{eqn:phiPlus}
    	\\
    	&	\Psi_{-1}(\xi_{\alpha},\zeta_{\alpha\alpha}) = \frac{-\xi_{\alpha} + \zeta_{\alpha\alpha}}{2},
    	\label{eqn:phis}
    \end{align}
and considers a product-form associated with the discrete velocities $\bm{c}_i$ (\ref{eq:d3q27vel}),
    \begin{equation}\label{eq:prod}
    	\Psi_i= \Psi_{c_{ix}}(\xi_x,\zeta_{xx}) \Psi_{c_{iy}}(\xi_y,\zeta_{yy}) \Psi_{c_{iz}}(\xi_z,\zeta_{zz}).
    \end{equation}
All pertinent populations below are determined by specifying the parameters $\xi_\alpha$ and $\zeta_{\alpha\alpha}$ in the product-form (\ref{eq:prod}). 
The two-dimensional version of the model on the D2Q9 lattice is obtained by omitting the $z$-component in all formulas. After matching moments with their continuous counter-parts the parameters are set as,
    \begin{align}
    &\xi_{\alpha}=u_{\alpha},\\
    &\zeta_{\alpha\alpha}=c_s^2+u_{\alpha}^2,\label{eq:prod_form_xi}
    \end{align} 
and the local equilibrium populations are represented with the product-form \eqref{eq:prod},
    \begin{equation}\label{eq:LBMeq}
        f_i^{\rm eq}=
        \rho\prod_{\alpha=x,y,z}\Psi_{c_{i\alpha}}\left(u_\alpha,c_s^2+u_{\alpha}^2\right).
    \end{equation}
This form of the discrete equilibrium populations, when $c_s^2=\bar{r}_0 T_0/3$ is equivalent to third-order quadrature-based scheme with a full expansion of the distribution function.
\paragraph{Alternative to polynomial equilibria: Entropic equilibria}
As an alternative to the classical discrete equilibrium construction approach where all degrees of freedom are used to fulfill moments constraints, the entropic approach adds minimization of an entropy functional to the list of constraints, changing the equilibrium construction problem into a constraint minimization problem. While a number of different discrete entropy functions have been proposed in the literature the most commonly used one is:
    \begin{equation}
    H_{w_i,c_i} = \sum_{i=1}^{Q} f_i \ln\left(\frac{f_i}{w_i}\right).
    \end{equation}
Minimization of this functional under constraints on moments of order zero and one, leads to the following well-known entropic equilibrium:
    \begin{equation}\label{eq:Entropic_isothermal_EDF}
    f_i^{\rm eq} =  w_i \rho  \prod_{\alpha=x,y} \left(2-\sqrt{{u_\alpha}^2/c_s^2 + 1}\right) {\left(\frac{2u_\alpha + \sqrt{{u_\alpha}^2/c_s^2 + 1}}{1-u_\alpha}\right)}^{c_{i,\alpha}}.
    \end{equation}
One of the most interesting feature of this equilibrium, contrary to polynomial equilibria is that, as demonstrated in \cite{hosseini_lattice_2023,hosseini_entropic_2023}, it guarantees unconditional linear stability.
\subsubsection{Lattice Boltzmann equations}
\paragraph{From discrete velocity Boltzmann to the lattice Boltzmann method}
To go to the final form of the lattice Boltzmann equations, two main ingredients are to be used: (a) integration of the discrete velocity Boltzmann equation along their \emph{constant} characteristics and (b) a re-definition of the discrete distribution functions. The former step results in:
\begin{equation}\label{eq:integrate_along_chars}
    f_i\left(\bm{x}+\bm{c}_i \delta t, t+\delta t\right) - f_i\left(\bm{x}, t\right) = \int_{t}^{t+\delta t} \Omega_i\left(\bm{x}(t'), t'\right)dt',
\end{equation}
where the term on the right-hand side, representing collision, has to be approximated via the trapezoidal rule, in order to keep the scheme second-order accurate:
\begin{equation}\label{eq:trapezoidal_rule}
    \int_{t}^{t+\delta t} \Omega_i\left(\bm{x}(t^{'}), t^{'}\right)dt^{'} = \frac{\delta t}{2}\Omega_i\left(\bm{x}, t\right) \\ +  \frac{\delta t}{2}\Omega_i\left(\bm{x}+\bm{c}_i\delta t, t+\delta t\right) + \mathcal{O}\left(\delta t^3\right).
\end{equation}
However, as observed here, application of the trapezoidal rule would make the scheme implicit and therefore not attractive regarding efficiency. The second ingredient, i.e. redefinition of the discrete distribution function as:
\begin{equation}\label{eq:redefined_f}
    \bar{f}_i = f_i - \frac{\delta t}{2}\Omega_i,
\end{equation}
changes the system of equations into a fully explicit scheme:
\begin{equation}\label{eq:explicit_lbm}
\bar{f}_i\left(\bm{x}+\bm{c}_i\delta t, t+\delta t\right) - \bar{f}_i\left(\bm{x}, t\right) = \frac{\delta t}{\bar{\tau}}\left( f^{\rm eq}_\alpha\left(\bm{x}, t\right) - \bar{f}_i\left(\bm{x}, t\right)\right),
\end{equation}
where $\bar{\tau}$ is now defined as:
\begin{equation}\label{eq:2-3-2-10}
\bar{\tau} = \tau + \delta t/2.
\end{equation}
\paragraph{The lattice Boltzmann method for incompressible flows}
A multi-scale analysis of the above-listed system of equations shows that it recovers the \emph{isothermal} continuity plus Navier-Stokes system of equations for an ideal equation of state at reference temperature:
\begin{equation}
    \bar{r}_0 T_0 = \frac{\delta x^2}{3\delta t^2},
\end{equation}
where the coefficient $1/3$ is specific to the third-order quadrature. Note that the recovered system of macroscopic equations further admits a defect in the viscous stress tensor; For the classical second-order polynomial expansion defects are present both in shear and bulk viscosity scaling in both cases as $\propto \mathcal{U}_\alpha^2 \delta t^2/\delta x^2$, where $\mathcal{U}$ is the characteristic velocity along the $\alpha$-axis. For the full polynomials expansion or product-form equilibria, this defect is only present in the bulk viscosity.\\
This means that under acoustic scaling, i.e. $\delta x/\delta t={\rm const}$ the solver converges to the \emph{compressible isothermal Navier-Stokes} equations, as for a fixed characteristic velocity $\mathcal{U}$ the Mach number remains constant in the limit of $\delta t\rightarrow 0$. Furthermore, under acoustic scaling the defects in the effective viscosities do not vanish. Under diffusive scaling on the other hand, i.e. $\delta x^2/\delta t = {\rm constant}$, the solver converges to the \emph{incompressible Navier-Stokes} equations, as $\mathcal{U}/c_s\rightarrow 0$ in the limit of $\delta t\rightarrow 0$ and the defect in the effective viscosities also goes to zero.

\section{Lattice Boltzmann models for compressible reacting flows}
We have now introduced the main ingredients of classical Lattice-Boltzmann methods, and shown that they allow to recover the continuity \eqref{eq:mass} and momentum \eqref{eq:momentum} equations of a weakly compressible gas at constant temperature $T_0$. This is indeed not sufficient for combustion applications, where energy \eqref{eq:energy} and species \eqref{eq:species} are absolutely required. 

This Section is divided into 3 main subsections. First, we shall explore different alternatives for the resolution of the additional equations (energy and species) in Section~\ref{subsec:energy_species}. Second, we will detail the required changes to the lattice Boltzmann formulation in Section~\ref{subsec:compressibleLBM}. Finally, we will list the reactive flow configurations successfully simulated by LBM solvers and discuss their performance in Section~\ref{subsec:wrapup}. 
\subsection{Energy and species balance equations}\label{subsec:energy_species}
\subsubsection{Double distribution function lattice Boltzmann approach for thermal flows}
\begin{figure*}
    \centering	    
    \includegraphics[width=12cm,keepaspectratio]{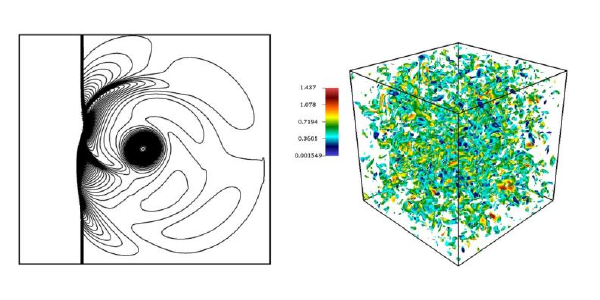}
    \caption{Illustration of applications of the model of Eq.~\eqref{eq:g_pop_lbm_coll_stream}. (Left) Sound pressure field for shock–vortex interaction with advection Mach number of $1.2$ and vortex Mach number set to $0.25$ at $t^*=6$. (Right) Iso-surface of velocity divergence colored by local Mach number for compressible decaying turbulence at Ma$=0.5$ and $t^*=0.4$. Images are reproduced from \cite{saadat_extended_2021}.}
    \label{Fig:PECS_ddf_gpop_applications}
\end{figure*}

\paragraph{Kinetic models}
Historically, the starting point of double distribution function (DDF) approaches is rooted in the need for simulations with variable Prandtl numbers and specific heat capacities, as alternatives to Holway's ellipsoidal statistics~\cite{holway_kinetic_1965} or Shakhov's model~\cite{shakhov_generalization_1972}. This is usually achieved by introducing a second distribution function $g$ to carry a form of energy following~\cite{rykov_model_1976}, which is not uniquely defined. The earliest occurrence of a double distribution function approach is documented in \cite{he_novel_1998} where the authors introduced:
\begin{equation}
    g(\bm{v},\bm{x},t) = \frac{{\left(v_\alpha-u_\alpha\right)}^2}{2} f(\bm{v},\bm{x},t).
\end{equation}
Multiplying the Boltzmann equation, i.e. the balance law for $f$ by the coefficients in the definition of the $g$-distribution function one obtains the balance law for the latter:
\begin{equation}
    \frac{\partial g}{\partial t} + v_\alpha \frac{\partial g}{\partial x_\alpha} = \frac{1}{\tau_g}\left(g^{\rm eq} - g\right) + f q,
\end{equation}
where the additional non-homogeneous contribution $q$ is:
\begin{equation}
    q = \left(u_\alpha - v_\alpha\right) \left[ \partial_t u_\alpha + v_\beta\frac{\partial u_\alpha}{\partial x_\beta}\right].
\end{equation}
In this model, the total energy $E$ is computed as:
\begin{equation}
    \rho E = \int_{\bm{v}} \left[\frac{\bm{u}^2}{2}f(\bm{v},\bm{x},t) + g(\bm{v},\bm{x},t) \right] d\bm{v}.
\end{equation}
Some comments on this approach are necessary:
\begin{itemize}
    \item This model, through the choice of parameter $\tau_g$ allows for a variable Prandtl number Pr.
    \item The model assumes a mono-atomic molecule as no degrees of freedom in addition to translational are taken into account.
    \item The model involves space and time derivatives of macroscopic fields.
\end{itemize}
To alleviate the last issue, Guo et al. proposed to carry total energy with $g$ instead~\cite{guo_thermal_2007}:
\begin{equation}
    g(\bm{v},\bm{x},t) = \frac{v_\alpha^2}{2} f(\bm{v},\bm{x},t).
\end{equation}
While this choice of a second distribution function leads to a much simpler balance law for $g$ it also comes with a limitation of the Prandtl number. Contrary to the previous choice of $g$ carrying internal energy where one could easily vary Pr by changing $\tau_g$, here the relaxation time in the collision operator controls both relaxation of internal and kinetic energy, therefore also affecting viscous heating. To allow for variable Pr, the authors proposed to decompose the collision term into kinetic and internal contributions, leading to the following balance law:
\begin{equation}\label{eq:g_pop_guo_balance}
    \frac{\partial g}{\partial t} + v_\alpha \frac{\partial g}{\partial x_\alpha} = \frac{1}{\tau_g} \left(g^{\rm eq} - g\right) + \frac{Z}{\tau_{gf}} \left(f^{\rm eq} - f\right),
\end{equation}
where
\begin{equation}
    Z = \frac{v_\alpha^2}{2} - \frac{{\left(v_\alpha - u_\alpha\right)}^2}{2},
\end{equation}
and
\begin{equation}
    \frac{1}{\tau_{gf}} = \frac{1}{\tau_g} - \frac{1}{\tau}.
\end{equation}
Since $g$ carries the total energy it is computed solely as its zeroth-order moment:
\begin{equation}\label{eq_E_zeroth}
    \rho E = \int_{\bm{v}} g(\bm{v},\bm{x},t) d\bm{v}.
\end{equation}
In the same contribution the authors proposed a more generalized framework allowing to incorporate additional non-translational degrees of freedom into the model by defining $g$ as:
\begin{equation}
    g(\bm{v},\bm{x},t) = \frac{v_\alpha^2 + \eta_\beta^2}{2} f(\bm{v},\bm{x},t),
\end{equation}
where $\bm{\eta}$ is a vector with $\delta$ component, with $\delta$ the number of additional degrees of freedom, and summation over both $\alpha$ and $\beta$  is assumed. In this model the equilibrium distribution function is:
\begin{equation}
    f^{\rm eq}(\bm{v},\bm{\eta},\bm{x},t) = \rho {\left(2\pi r T\right)}^{(\delta + D)/2} \exp\{-\frac{{(\bm{v}-\bm{u})}^2 + \bm{\eta}^2}{2 r T}\},
\end{equation}
and the total energy is computed as:
\begin{equation}
    \rho E = \int_{\bm{v}} \int_{\bm{\eta}} g(\bm{v}, \bm{\eta}, \bm{x},t) d\bm{\eta} d\bm{v}.
\end{equation}
While Guo et al. originally proposed this decoupling for the low Mach limit, it was extended to compressible regimes in \cite{li_coupled_2007} where the authors used a thirteen-velocity lattice for the $f$ distribution function to eliminate the deviations in the third-order moments of the equilibrium distribution function. A realization of this model on the standard third-order quadrature-based lattice was proposed in \cite{li_coupling_2012}. The approach originally proposed in~\cite{guo_thermal_2007} has been routinely used since then in a wide number of publications for both compressible and incompressible flows, see for instance~\cite{meng_lattice_2008,feng_three_2015,feng_compressible_2016,liu_coupled_2018}. Another realization of the double distribution function method relying on internal energy was also proposed in \cite{karlin_consistent_2013}. As noted by the authors, re-writing the balance equation of Eq.~\eqref{eq:g_pop_guo_balance} as:
\begin{equation}
    \frac{\partial g}{\partial t} + v_\alpha \frac{\partial g}{\partial x_\alpha} = \frac{1}{\tau} \left(g^{\rm eq} - g\right) + \frac{1}{\tau_{gf}} \left(g^* - g\right),
\end{equation}
in the case of the model in \cite{guo_thermal_2007}:
\begin{equation}
    g^* = g^{\rm eq} + Z\left(f-f^{\rm eq}\right),
\end{equation}
while for \cite{karlin_consistent_2013}:
\begin{equation}
    g^* = g^{\rm eq} + 2 v_\alpha u_\beta \left( \Pi_{\alpha\beta} - \Pi^{\rm eq}_{\alpha\beta}\right).
\end{equation}
Note that both realizations lead to the same hydrodynamic equation and in the case of the third-order quadrature-based lattices, even to the same discrete equations~\cite{karlin_consistent_2013}. This realization has also been used for a variety of compressible flow simulations, see for instance \cite{saadat_lattice_2019,saadat_extended_2021}.\\
\paragraph{Lattice Boltzmann equations}
Discretization in the space of particles velocities $\bm{v}$ proceeds very similarly to that of the probability distribution function $f$, either through projection onto the space of Hermite polynomials or via the product form construction. In the product form approach discussed in \cite{karlin_consistent_2013,saadat_extended_2021} similar to Eq.~\eqref{eq:LBMeq}:
\begin{equation}\label{eq:LBMeq_product_g}
    g_i^{\rm eq}=
    \rho\prod_{\alpha=x,y,z}\Psi_{c_{i\alpha}}\left(\mathcal{O}_\alpha, \mathcal{O}_\alpha^2\right)E,
\end{equation}
where the operator $\mathcal{O}_\alpha$ acts on any smooth function $A(\bar{r} T, u_\alpha)$ as:
\begin{equation}
    \mathcal{O}_\alpha A = \bar{r} T \frac{\partial A}{\partial u_\alpha} + u_\alpha A.
\end{equation}
The discrete-in-particles speed-space can then be integrated along characteristics to obtain the corresponding collision-streaming equations:
\begin{equation}\label{eq:g_pop_lbm_coll_stream}
    \bar{g}_i(\bm{x}+\bm{c}_i\delta t, t+\delta t) - \bar{g}_i(\bm{x}, t) = \frac{\delta t}{\bar{\tau}}\left( g^{\rm eq}_i(\bm{x}, t) - \bar{g}_i(\bm{x}, t) \right)\\  + \frac{\delta t}{\bar{\tau}_{gf}} \left( g^*_i(\bm{x}, t) - \bar{g}_i(\bm{x}, t) \right),
\end{equation}
where the new distribution function is:
\begin{equation}
    \bar{g}_i = g_i - \frac{\delta t}{2}\left[ \frac{1}{\tau}\left(g_i^{\rm eq} - g_i\right) + \frac{1}{\tau_{gf}}\left( g^*_i - g_i \right) \right]
\end{equation}
and
\begin{equation}
    \frac{\delta t}{\bar{\tau}_{gf}} = \frac{\delta t}{\bar{\tau}_{g}} - \frac{\delta t}{\bar{\tau}},
\end{equation}
with
\begin{equation}
    \frac{\bar{\tau}_g}{\delta t} = \frac{\lambda}{p C_v} + \frac{1}{2}.
\end{equation}
The total energy can then be obtained by summing discrete distribution functions:
\begin{equation}
    \rho E = \sum_{i=0}^{Q-1} g_i.
\end{equation}
A multi-scale analysis shows that the models above, at the Euler level, recover:
\begin{equation}
    \frac{\partial\rho E}{\partial t^{(1)}} + \frac{\partial \rho u_\alpha H }{\partial x_\alpha} + \frac{\partial \rho u_\alpha u_\beta^2/2}{\partial x_\alpha} = 0,
\end{equation}
where $H$ is the enthalpy. At the Navier-Stokes level:
\begin{equation}
    \frac{\partial\rho E}{\partial t^{(2)}} + \frac{\partial \Pi_{\alpha}(g_i^{(1)})}{\partial x_\alpha} = 0,
\end{equation}
with:
\begin{equation}\label{eq:noneq_g_flux}
    \Pi_{\alpha}(g_i^{(1)}) = -\left(\frac{\bar{\tau}_g}{\delta t}-\frac{1}{2}\right) p\frac{\partial H}{\partial x_\alpha} + u_\beta\Pi_{\alpha\beta}(f_i^{(1)}),
\end{equation}
where the first term is the Fourier diffusive flux while the second term is viscous heating. The double distribution function approach in combination with a proper lattice Boltzmann solver for density and momentum balance, detailed in next sections, has been used to model trans- and supersonic flows, for instance in \cite{saadat_extended_2021}. A few results are illustrated in Fig.~\ref{Fig:PECS_ddf_gpop_applications}.
\paragraph{Multi-species flows} For the extension of this model to the case of multi-species flows in the context of a mixture-averaged formulation a few points must be noted:
\begin{itemize}
    \item In all models discussed here, at the $\epsilon^2$ order a Fourier flux of this form is recovered, see Eq.~\eqref{eq:noneq_g_flux}:
    \begin{equation}
        -\left(\frac{\bar{\tau}_g}{\delta t}-\frac{1}{2}\right) p\frac{\partial H}{\partial x_\alpha} = -\lambda \frac{\partial T}{\partial x_\alpha},
    \end{equation}
    which holds if enthalpy is only function of temperature. For a mixture-averaged formulation with multiple species, $H = H(T, Y_k)$, which would lead to:
    \begin{equation}
        -\left(\frac{\bar{\tau}_g}{\delta t}-\frac{1}{2}\right) p\frac{\partial H}{\partial x_\alpha} = -\lambda \frac{\partial T}{\partial x_\alpha} \\ - \left(\frac{\bar{\tau}_g}{\delta t}-\frac{1}{2}\right) p \sum_{k=0}^{N_{sp}-1}  H_k \frac{\partial Y_k}{\partial x_\alpha},
    \end{equation}
    where $H_k$ is the enthalpy of the $k^{\rm th}$ species.
    \item In multi-species flows, diffusive mass flux leads to a net transport of enthalpy which is absent in single-component flows. 
\end{itemize}
A solution to both these shortcomings was proposed in~\cite{sawant_consistent_2021} to recover consistent hydrodynamics, where the pseudo-equilibrium $g^{*}_i$ is amended with two additional terms, one neutralizing the error in the Fourier diffusion and one introducing enthalpy flux through mass diffusion:
\begin{equation}
    g^{*}_i = g_i^{\rm eq} + \frac{2w_i}{c_s^2} c_{i\alpha} \left[u_\beta\left(\Pi_{\alpha\beta} - \Pi^{\rm eq}_{\alpha\beta}\right) \right. \\ \left. + p \sum_{k=0}^{N_{sp}-1}  H_k \frac{\partial Y_k}{\partial x_\alpha} + \rho \sum_{k=0}^{N_{sp}-1}  Y_k H_k V_{k\alpha} \right].
\end{equation}
\subsubsection{Kinetic models for species balance equations}
\begin{figure}
    \centering
    \includegraphics[width=8cm,keepaspectratio]{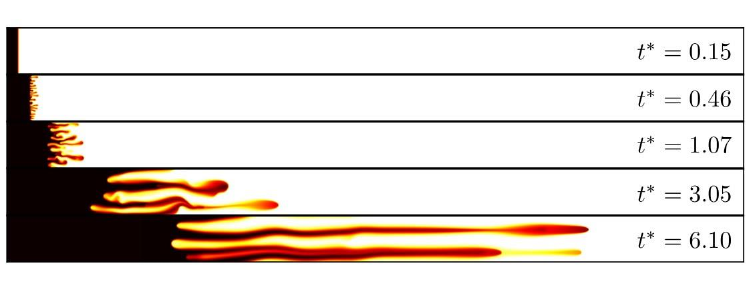}
    \caption{Illustration of application of multi-species model of \cite{vienne_lattice_2019}. Evolution of viscous fingering instability for a system with two species. Image reproduced from \cite{vienne_lattice_2021}. }
    \label{Fig:PECS_species_vienne_app}
\end{figure}

\begin{figure*}
    \centering
    \includegraphics[width=14cm,keepaspectratio]{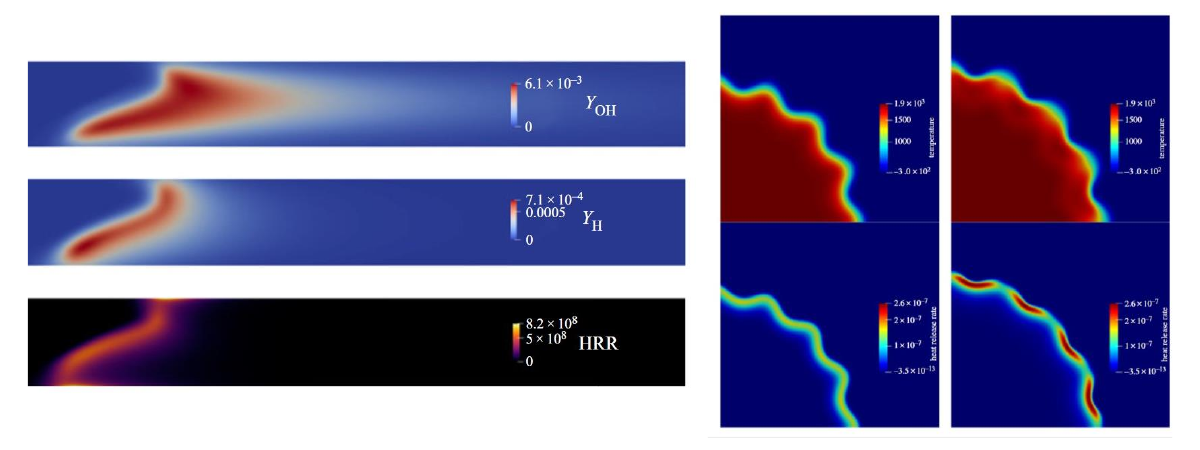}
    \caption{Illustration of applications of multi-species model of \cite{sawant_consistent_2021,sawant_consistent_2022}. (Left) Upper asymmetric hydrogen flame in micro-channel. (Right) Thermo-diffusive instability in radially expanding hydrogen flame. Images reproduced from \cite{sawant_consistent_2022,sawant_lattice_2021}. }
    \label{Fig:PECS_sawant_applications}
\end{figure*}

Over the past decades and starting in the early 2000's~\cite{luo_lattice_2002,luo_theory_2003} various attempt at developing lattice Boltzmann-based models for mixtures have been documented, see for instance~\cite{asinari_viscous_2005,asinari_consistent_2008,asinari_multiple-relaxation-time_2008,bennett_lattice_2012,chai_maxwell-stefan-theory-based_2019}. Some of these models are reviewed in this section.
\paragraph{Thermal mixture-averaged model of Kang et al}
In \cite{kang_thermal_2014}, the authors proposed a multi-component thermal model for catalytic systems. The model is an extension on previous work documented in  \cite{arcidiacono_simulation_2006,arcidiacono_lattice_2007,arcidiacono_lattice_2008,kang_lattice_2013}. It consists of $N_{sp}$ sets of lattice Boltzmann solvers, i.e. one per species:
\begin{equation}\label{eq:kang_model_lbm}
    g_{ki}(\bm{x}+\bm{c}_{ki}\delta t, t+\delta t) - g_{ki}(\bm{x}, t) = \frac{\delta t}{\bar{\tau}_{k1}}\left( g^{*}_{ki}(\rho_k, \bm{u}_k) - g_{ki}(\bm{x}, t)\right) + \frac{\delta t}{\bar{\tau}_{k2}}\left( g^{\rm eq}_{ki}(\rho_k, \bm{u}) - g_{ki}(\bm{x}, t)\right) + \psi_{ki}.
\end{equation}
The first point to note in this model is that post-streaming discrete distribution functions migrate to $\bm{x}+\bm{c}_{ki}\delta t$ meaning each species' lattice have different discrete velocity sizes. As discussed in previous sections, in lattice Boltzmann time-step, grid-size and reference temperature are tied through:
\begin{equation}
    \frac{\delta x^2}{\delta t^2} = \frac{\mathcal{R} T_0}{W},
\end{equation}
which in the context of this model where $W = W_k$ is different for each species, and assuming that the time-step size is the same for all solvers, would mean:
\begin{equation}
    \|c_{ki\alpha}\| = \frac{\delta x_k}{\delta t} = \sqrt{\frac{\mathcal{R} T_0}{W_k}},
\end{equation}
i.e. not all species will propagate on lattice. To overcome this issue, and following \cite{mccracken_lattice_2005} the authors proposed to set the grid-size to that needed for the lightest species in the system, and for other species to use interpolation in order to reconstruct distribution functions on the grid. The equilibrium, $g^{\rm eq}_{ki}(\rho_k, \bm{u})$ follows the product-form equilibrium of Eq.~\eqref{eq:LBMeq} with a few differences, namely:
\begin{align}
    &\xi_{k\alpha}=u_{\alpha} \sqrt{W_k},\\
    &\zeta_{k\alpha\alpha}= T + W_k u_{\alpha}^2,
\end{align} 
while for the pseudo-equilibrium $g^{*}_{ki}(\rho_k, \bm{u}_k)$:
\begin{align}
    &\xi_{k\alpha}=u_{k\alpha} \sqrt{W_k},\\
    &\zeta_{k\alpha\alpha}= T + W_k u_{k\alpha}^2.
\end{align} 
In this model the second relaxation time, $\bar{\tau}_{k2}$ sets the diffusivity to:
\begin{equation}
    \frac{\bar{\tau}_{k2}}{\delta t} = \frac{\rho_k D_k}{p_k} + \frac{1}{2},
\end{equation}
where $D_k$ is the mixture-average diffusion coefficient. The viscosity is set through the first relaxation time and using Wilke's formula:
\begin{equation}
    \frac{\bar{\tau}_{k1}}{\delta t} = \frac{\mu_k}{p\sum_{k'=0}^{N_{sp}-1} X_{k'} \phi_{kk'}} + \frac{1}{2},
\end{equation}
with
\begin{equation}
    \phi_{kk'} = \frac{1}{\sqrt{8}}\frac{1}{\sqrt{1+ \frac{W_k}{W_{k'}}}}{\left[ 1 + \sqrt{\frac{\mu_k}{\mu_{k'}}}{\left(\frac{W_{k'}}{W_k}\right)}^{1/4} \right]}^2.
\end{equation}
The term $\psi_{ki}$ in Eq.~\eqref{eq:kang_model_lbm} is a correction term accounting for: (a) a correction velocity ensuring that the global diffusive mass flux is null, and (b) corrections for equilibrium moments of order three and four not recovered by the first-neighbour lattice. The latter terms allow the scheme to recover the proper viscous stress tensor and non-equilibrium heat flux.
In this model the macroscopic properties are computed as:
\begin{align}
    &\rho_k=\sum_{i=0}^{Q-1} g_{ki},\\
    &\rho_k u_{k\alpha}=\sum_{i=0}^{Q-1} c_{ki\alpha} g_{ki},\\
    &\rho_k E_k=\sum_{i=0}^{Q-1} c_{ki\alpha}^2 g_{ki}.
\end{align}
A multi-scale expansion of the model shows that it recovers the mixture-averaged multi-species equations and the Hirschfelder-Curtiss approximation with the mass corrector.
\paragraph{Force-based approach of Vienne et al}
In \cite{vienne_lattice_2019}, following the kinetic model of \cite{kerkhof_toward_2005}, Vienne et al. proposed a lattice Boltzmann model for isothermal multi-species mixtures recovering the Maxwell-Stefan system of equations. Considering a mixture made up of $N_{sp}$ individual species, they proposed a coupled system of $N_{sp}$ lattice Boltzmann solvers:
\begin{equation}
    g_{ki}(\bm{x}+\bm{c}_i\delta t, t+\delta t) - g_{ki}(\bm{x}, t) =  \frac{\delta t}{\bar{\tau}_k}\left( g^{\rm eq}_{ki}(\rho_k, \bm{u}_k) - g_{ki}(\bm{x}, t)\right) + \mathcal{S}_i,
\end{equation}
where $\mathcal{S}_i$ is here to introduce external body forces, realized using Guo's approach~\cite{guo_discrete_2002}:
\begin{equation}
    \mathcal{S}_i = \left( 1-\frac{\delta t}{2\bar{\tau}_k}\right) w_i \left( \frac{c_{i\alpha} - u_{k\alpha}}{c_s^2} + \frac{(c_{i\beta}u_{k\beta})c_{i\alpha}}{c_s^4}\right)F_{k\alpha},
\end{equation}
where $\bm{F}_k$ represents the body force. In this model
\begin{equation}
    \rho_k = \sum_{i=0}^{Q-1} g_{ki},
\end{equation}
and 
\begin{equation}\label{eq:vienne_mom_def}
    \rho_k u_{k\alpha} = \sum_{i=0}^{Q-1} c_{i\alpha}g_{ki} + F_{k\alpha}.
\end{equation}
The interaction between different species driving diffusion is introduced via a body force defined as:
\begin{equation}\label{eq:vienne_force}
    F_{k\alpha} = -p\sum_{k'=0}^{N_{sp}-1}\frac{X_k X_{k'}}{\mathcal{D}_{kk'}}(u_{k\alpha}-u_{k'\alpha}),
\end{equation}
where $\mathcal{D}_{kk'}$ represents the binary diffusion coefficients. As noted by the authors, the circular inter-dependence between the force of Eq.~\eqref{eq:vienne_force} and the momenta of individual species of Eq.~\eqref{eq:vienne_mom_def} make the scheme implicit. A multi-scale analysis shows that this model recovers the following multi-component isothermal mass
\begin{equation}
    \frac{\partial \rho_k}{\partial t} + \frac{\partial \rho_k u_{k\alpha}}{\partial x_\alpha} = 0,
\end{equation}
and momentum balance equations
 \begin{equation}
     \frac{\partial \rho_k u_{k\alpha}}{\partial t} + \frac{\partial \rho_k u_{k\alpha}u_{k\beta}}{\partial x_\beta} + \frac{\partial p_k}{\partial x_\alpha}  - \frac{\partial }{\partial x_\beta}\left(\mu_k\frac{\partial u_{k\beta}}{\partial x_\alpha}+\mu_k\frac{\partial u_{k\alpha}}{\partial x_\beta}\right)  + p \sum_{k'=0}^{N_{sp}-1}\frac{X_k X_{k'}}{\mathcal{D}_{kk'}}(u_{k\alpha}-u_{k'\alpha}) = 0,
 \end{equation}
where the bulk viscosities $\mu_k$ are defined as:
 \begin{equation}
     \frac{\bar{\tau}_k}{\delta t} = \frac{\mu_k}{\rho_k c_s^2} + \frac{1}{2}.
 \end{equation}
The model has been successfully used to study miscible multi-component flow behaviors such as viscous fingering, see Fig.~\ref{Fig:PECS_species_vienne_app}.
An extension of this model to thermal and reacting cases is yet to done.
\paragraph{Model of Sawant et al} 
In \cite{sawant_consistent_2021,sawant_consistent_2022}, the authors proposed a kinetic model to recover the Stefan--Maxwell diffusion model. Each component is described by a set of populations $g_{ki}$. The discrete-velocity time evolution equation is,
\begin{equation}
\frac{\partial g_{ki}}{\partial t} + c_{i\alpha} \frac{\partial g_{ki}}{\partial x_\alpha} = \sum_{k'\neq k} \frac{1}{\theta_{kk'}} \left[ \left( \frac{g_{ki}^{\rm eq}-g_{ki}}{\rho_k} \right) - \left( \frac{f_{k'i}^{\rm eq}-f^*_{k'i}}{\rho_{k'}} \right) \right].
\label{ch1-eqn:stefanMaxwell} 
\end{equation}
The species densities are computed as zeroth-order moment of the discrete distribution functions:
\begin{equation}
\rho_k=\sum_{i=0}^{Q-1}g_{ki}.
\label{ch1-eq:density1}
\end{equation}
The symmetric set of relaxation times $\theta_{kk'}=\theta_{k'k}$ is related to the binary diffusion coefficients.
The first-order moments of the distribution functions are,%
\begin{equation}
\rho_k u_{k\alpha}=\sum_{i=0}^{Q-1} g_{ki}c_{i\alpha}.
\label{ch1-eq:ua2}
\end{equation}
The quasi-equilibrium populations $g_{ki}^*$ satisfy the following constraints on moments,  
\begin{align}
\sum_{i=0}^{Q-1} g_{ki}^{*} &=  \rho_k,
\label{ch1-eq:rhoqe}\\
\sum_{i=0}^{Q-1} g_{ki}^{*} c_{i\alpha} &=  \rho_k u_{k\alpha}.
\label{ch1-eq:ua}
\end{align}
The momenta of the individual species sum up to the mixture momentum,
\begin{equation}
\sum_{k=0}^{N_{sp}-1} \rho_k u_{k\alpha} = \rho u_\alpha.
\label{ch1-eqn:mixtureMomentum}
\end{equation}
The equilibrium populations $g_{ki}^{\rm eq}$ are subject to the following constraints:
\begin{align}
\sum_{i=0}^{Q-1} g_{ki}^{\rm eq} &=  \rho_k,
\label{ch1-eqn:f0ZerothMoment}\\
\sum_{i=0}^{Q-1} g_{ki}^{\rm eq} c_{i\alpha} &=  \rho_k u_\alpha,
\label{ch1-eq:u}
\\
\sum_{i=0}^{Q-1} g_{ki}^{\rm eq} c_{i\alpha} c_{i\beta}&=  p_k\delta_{\alpha\beta} + \rho_k u_\alpha u_\beta.
\label{ch1-eq:Pa}
\end{align}
In Eq.~\eqref{ch1-eq:Pa}, the partial pressure $p_k$ depends on the mixture temperature $T$ which is obtained from the energy balance lattice Boltzmann solver. Noting that
\begin{equation}
    \theta_{kk'} = \frac{\mathcal{D}_{kk'}}{p X_k X_{k'}},
\end{equation}
and using the equation of state the kinetic model can be re-written as:
\begin{equation}
\frac{\partial g_{ki}}{\partial t} + c_{i\alpha}\frac{\partial g_{ki}}{\partial x_\alpha} =\\ 
\sum_{k'\neq k} \left(\frac{ \bar{W} \mathcal{R} T }{W_k W_{k'} \mathcal{D}_{kk'}}\right) 
\left[Y_{k'} \left( g_{ki}^{\rm eq}-g_{ki} \right) - Y_k \left( g_{k'i}^{\rm eq}-g^*_{k'i}\right)\right].
\label{ch1-eqn:stefanMaxwellNumericalStable} 
\end{equation}
This equation, for the sake of convenience, is recast in the form of a relaxation equation:
\begin{equation}
    \frac{\partial g_{ki}}{\partial t} + c_{i\alpha}\frac{\partial g_{ki}}{\partial x_\alpha} = \frac{1}{\tau_k}\left(g^{m eq}_{ki} - g_{ki}\right) - F_{ki},
\end{equation}
where
\begin{equation}
\frac{1}{\tau_k} = \sum_{k'\ne k} \frac{Y_{k'}}{\tau_{kk'}} = r_k T\left(\sum_{k'\neq k} \frac{X_{k'}}{\mathcal{D}_{kk'}}\right),
\label{ch1-eqn:tau}
\end{equation}
and
\begin{equation}
F_{ki} = Y_k \sum_{k'\neq k} \frac{1}{\tau_{kk'}}  \left( g_{k'i}^{\rm eq}-g_{k'i}^* \right).
\label{ch1-eqn:fStar}
\end{equation}
This form of the equation can then be integrated along characteristics to obtain the lattice Boltzmann equation. The model recovers the compressible mixture-averaged multi-species equation with the Maxwell-Stefan velocity for species diffusion. It has been successfully used for a variety of cases involving combustion applications with detailed chemistry, as illustrated in Fig.~\ref{Fig:PECS_sawant_applications}.

\subsubsection{Passive-scalar lattice Boltzmann models}
So-called passive-scalar lattice Boltzmann solvers, are models where only conservation of the zeroth-order moment of the distribution function is ensured by the collision operator. In such models, to solve an advection-diffusion-reaction partial differential equation for a field $\Psi$, a distribution function $g_i$ is defined such that:
\begin{equation}
    \sum_{i=0}^{Q-1}g_i + \frac{\delta t}{2} S = \Psi,
\end{equation}
where $S$ is the source term. A classical non-homogeneous collision-streaming equation of the form:
\begin{equation}\label{eq:lbm_passive_scalar}
    g_i(\bm{x}+\bm{c}_i \delta t, t+\delta t) - g_i(\bm{x}, t) = -\frac{\delta t}{\bar{\tau}}\left[ g_i^{\rm eq}(\Psi, \bm{u} ) - g_i(\bm{x}, t) \right]  + (1-\frac{\delta t}{2\bar{\tau}}) \mathcal{S}_i,
\end{equation}
with
\begin{equation}
    \sum_{i=0}^{Q-1} \mathcal{S}_i = S,
\end{equation}
and
\begin{equation}\label{eq:passive_equilibrium}
    g_i^{\rm eq}(\Psi, \bm{u} ) = \frac{\Psi f_i^{\rm eq}(\Psi, \bm{u} )}{\rho},
\end{equation}
leads to a macroscopic equation of the form:
\begin{equation}\label{eq:passive_scalar_macro_limit}
    \frac{\partial \Psi}{\partial t} + \frac{\partial \Psi u_\alpha}{\partial x_\alpha} \\ + \frac{\partial }{\partial x_\alpha}\left(\frac{1}{2} - \frac{\bar{\tau}}{\delta t}\right) \left(\frac{\partial \Psi u_\alpha}{\partial t} + \frac{\partial \Pi_{\alpha\alpha}(g_i^{\rm eq})}{\partial x_\alpha}\right) = S.
\end{equation}
Note that in the literature, Eq.~\eqref{eq:passive_equilibrium} has been used both with first and second-order polynomials expansions. Depending on the choice of the order of expansion the diffusion term will admit errors of different forms. For instance, for a linear equilibrium,
\begin{equation}
    \frac{\partial \Psi u_\alpha}{\partial t} + \frac{\partial \Pi_{\alpha\alpha}(g_i^{\rm eq})}{\partial x_\alpha} = c_s^2\frac{\partial \Psi}{\partial x_\alpha} + \frac{\partial \Psi u_\alpha}{\partial t}.
\end{equation}
On that note, let us now discuss the passive scalar approach in the specific context of the species mass and energy balance equations. Although a number of different works have presented modified passive scalar approaches for non-linear dependence of the diffusion driving force on the zeroth-order moment even in the context of multi-species flows (see for instance~\cite{hosseini_mass-conserving_2018,hosseini_weakly_2020}), here we will limit our discussion to models that have been used for combustion simulation.
\paragraph{Energy balance equation}
The energy balance equation can be written in a variety of different ways, see \cite{poinsot_theoretical_2005}. Here we will only discuss the form involving temperature:
\begin{equation}
    \frac{\partial T}{\partial t} + u_\alpha\frac{\partial T}{\partial x_\alpha} - \frac{1}{\rho \bar{c}_p}\frac{\partial}{\partial x_\alpha}\left(\lambda \frac{\partial T }{\partial x_\alpha}\right) + \frac{\dot{\omega}_T}{\rho \bar{c}_p} = 0. 
\end{equation}
The classical approach to recover this balance equation is to set $\Psi = T$ which would lead to the following shortcomings:
\begin{itemize}
    \item An error of the form $T\partial u_\alpha/\partial x_\alpha$ in the convection term.
    \item An error of the form:
    \begin{equation*}
        -\frac{\lambda}{\rho \bar{c}_p} \frac{\partial T}{\partial x_\alpha} \frac{\partial \rho \bar{c}_p}{\partial x_\alpha},
    \end{equation*}
    in the Fourier diffusion term.
    \item The enthalpy flux due to species mass diffusion is missing.
\end{itemize}
While one can overcome these issues via alternative forms of the equilibrium distribution function (see for instance~\cite{hosseini_lattice_2019}), the simplest way to circumvent these issues is to define the source term $S$ in Eq.~\eqref{eq:passive_scalar_macro_limit} as:
\begin{equation}
    S = -\frac{\dot{\omega}_T}{\rho \bar{c}_p} + T\frac{\partial u_\alpha}{\partial x_\alpha} - \frac{\lambda}{\rho \bar{c}_p}\frac{\partial T}{\partial x_\alpha}\frac{\partial \rho \bar{c}_p}{\partial x_\alpha} - \sum_{k=0}^{N_{sp}-1} \frac{c_{pk}}{\bar{c}_p} Y_k V_{k\alpha}\frac{\partial T}{\partial x_\alpha}.
\end{equation}
Such an approach, among others, has been used in \cite{lei_study_2021,lei_pore-scale_2023}. Note that in \cite{lei_study_2021,lei_pore-scale_2023} the last term was not considered. A similar approach can be undertaken for the case were the enthalpy or energy balance equation is targeted.\\

\paragraph{Species mass balance equations}
For the sake of simplicity let us assume that the species mass fraction balance equation is targeted. Taking the zeroth-order moment of the distribution function to be mass fraction, $Y_k$ and neglecting for the time being leading-order errors in the multi-scale analysis one would recover a diffusion term of the form:
    \begin{equation*}
        -\frac{\partial}{\partial x_\alpha} \left[ \left(\frac{1}{2} - \frac{\bar{\tau}}{\delta t}\right) \frac{\partial Y_k}{\partial x_\alpha}\right],
    \end{equation*}
which using $\bar{\tau}/\delta t=D_k/c_s^2 + 1/2$ recovers the generalized Fick approximation. With that the passive scalar approach is confronted with a number of issues:
\begin{itemize}
    \item This form of diffusion is only valid in the limit of vanishing density changes as in the non-conservative form of the balance equation there is a factor $1/\rho$ in front of the diffusion term.
    \item The form of the convection term as recovered in Eq.~\eqref{eq:passive_scalar_macro_limit} admits an error of the form $Y_k\partial u_\alpha/\partial x_\alpha$, which only vanishes for incompressible flows.
    \item It is well known that the generalized Fick approximation does not conserve overall mass, unless either $N_{sp} = 2$ or $D_k=D$, $\forall k$. To deal with that issue there are two approaches; If mass fraction of one particular species is dominant everywhere (e.g. that of ${\rm N}_2$ in combustion with air) the balance equation for that species is not explicitly solved and one sets $Y_{{\rm N}_2} = \sum_{k=0,k\neq {\rm N}_2}^{N_{sp}-1}$. A more general approach, valid also for non-dilute mixture is to introduce a mass corrector.
    \item In cases where the driving force of the diffusive flux is a linear function of the variable for which the balance equation is solved, for instance the Fick approximation, the passive scalar model can be used as is. However, for models where the driving force of diffusive flux is a non-linear function that depends on variables other than the zeroth-order moment, for instance the Hirschfelder-Curtiss approximation, the passive scalar approach would lead to errors of the same order as the diffusive flux itself.
\end{itemize}
A number of different approaches have been proposed in the literature to account for these shortcomings, see for instance~\cite{hosseini_mass-conserving_2018,hosseini_weakly_2020}. One of the most straightforward approaches, as used in \cite{lei_study_2021,lei_pore-scale_2023}, is to put all corrections into a source term. For instance, assuming one targets the mass fraction balance equation with the generalized Fick approximation, the source term $S$ would be:
\begin{equation}
    S = Y_k\frac{\partial u_\alpha}{\partial x_\alpha} + \frac{D_k}{\rho} \frac{\partial Y_k}{\partial x_\alpha} \frac{\partial \rho}{\partial x_\alpha}.
\end{equation}
Note that this approach as used in \cite{lei_study_2021,lei_pore-scale_2023} still comes short with respect to the mass corrector and the more appropriate diffusion velocity closure. A number of solutions to account for the mass corrector have been proposed in the literature, taking advantage of the non-equilibrium part of the distribution function, see for instance \cite{hosseini_mass-conserving_2018}.

\subsubsection{Hybrid models: Finite difference and finite volume solvers for energy and species}\label{subsec_hybrid}
Hybrid models are closest to multiple distribution functions approaches. In multiple distribution functions, let us remind that $\rho E$ (or an alternative energy form) and $\rho Y_k$ correspond to the zeroth order of separate distribution functions, see Eq.~\eqref{eq_E_zeroth}. 

When the number of species to be considered becomes large - typically $\mathcal{O}(10-100)$, the memory required to solve all scalars increases very quickly, which may become prohibitive for detailed chemistry descriptions.
Hybrid models reduce the memory load by introducing a single scalar for $(\rho E, \rho Y_k)$ instead of the number of discrete velocities. Each additional conserved scalar $\rho \phi$ (where $\phi$ may represent $E$ or $Y_k$) is solved by classical finite-difference or finite-volume (FD/FV) schemes, while continuity \eqref{eq:mass} and momentum \eqref{eq:momentum} equations are still solved via their associated distribution function $f_i$.

Let us now list the main advantages of the hybrid method: 
\begin{enumerate}
    \item The memory footprint is reduced, as only 1 additional scalar needs to be stored for each energy  or species equation, vs. 27 for a D3Q27 distribution.
    \item They are by construction free of Prandtl, Schmidt or $\gamma$ numbers limitations, since energy/species resolution is tackled separately.
    \item Since they use the same formalism as classical reactive flow solvers for energy and species equations, it is straightforward to take into account combustion-specific terms (turbulent combustion closures, advanced transport models, Soret effect or multi-component diffusion etc.), based on the experience accumulated over many decades using FD/FV solvers.
\end{enumerate}

In turn, hybrid methods suffer from the following drawbacks:
\begin{enumerate}
    \item Ensuring consistency between the LBM scheme (for continuity and momentum equations) and FD/FV schemes is not straightforward (at the opposite of, e.g. multi-speed or multiple distribution approaches). This can typically lead to disastrous spurious currents, as illustrated later in Fig.~\ref{fig:sf-velocomp}.
    \item FD/FV schemes based only on nearest-neighbor stencils (as used in most LBM solvers) are typically much more dissipative than LBM schemes \cite{marie2009comparison}.
\end{enumerate}

The first point is crucial in designing hybrid LBM schemes, and is therefore discussed at length hereafter. The impact of the second point is limited for most applications as long as the vortical and acoustic modes are left within the LBM part of the solver.

\paragraph{Which form to use for species/energy equations in a hybrid LBM scheme?}


Energy and species equations may be written under a large variety of forms (based on total energy, internal energy, temperature,...). While these forms are indeed equivalent for a continuous formulation, their coupling under discrete form with the LBM scheme may be very different.


Let us recall that for small perturbations and neglecting all dissipation terms (reducing to the multi-component Euler equations), the system (\ref{eq:mass}-\ref{eq:species}) may be linearized, and each perturbation can be decomposed into the so-called Kovasznay modes \cite{kovasznay1953,chukovasznay1958} (acoustic mode, 3 components of the vorticity modes, entropy mode, and 1 per species). 

For instance, the entropy mode of the Euler system follows the equation $$\frac{\partial s}{\partial t} + u_\alpha\frac{\partial s}{\partial x_\alpha}=0,$$ and is only weakly coupled with the rest of the system. 

For this reason, hybrid methods using an entropy equation were shown to provide reasonable results for moderately compressible flows \cite{farag2020pressure,farag2021bridge,coratger2021transonic} using several classical convective numerical schemes a priori unrelated with LBM: 
\begin{itemize}
    \item Second-order central difference schemes, potentially blended with upwind, \cite{tayyab2020experimental,tayyab2020hybrid,boivin2021tgv,tayyab2021volvo}
    \item Lax-Wendroff scheme \cite{wissocq2019extended,wissocq2022Etot}
    \item MUSCL schemes \cite{farag2020pressure,farag2021bridge,coratger2021transonic}
    \item Heun scheme \cite{wissocq2023deto},
    \item \ldots
\end{itemize}

Indeed, for reactive flows, the entropy equation is complex to derive in its general case. However, the enthalpy equation under non-conservative form
\begin{equation}\label{eq:energy_noncons}
    \rho\frac{\partial h}{\partial t} +  \rho u_i\frac{\partial h}{\partial x_i} = \frac{dP}{dt} - \frac{\partial q_i}{\partial x_i} +  \Pi_{ij}\frac{\partial u_i}{\partial x_j}
\end{equation}
is also a characteristic mode of the system - provided the pressure work $\frac{dP}{dt}$ is neglected, a very common assumption for low-Mach reactive flows.

Species also directly follow characteristic equations, provided they are written under non-conservative form
\begin{equation}\label{eq:species_noncons}
\rho\frac{\partial Y_k}{\partial t} + \rho u_i \frac{\partial Y_k}{\partial x_i} = \frac{\partial}{\partial x_i}(\rho Y_k V_{k,i}) + \dot{\omega}_k,    
\end{equation}

There is an alternative but not equivalent way of understanding how crucial this choice is. Consider the species equation in conservative form  
\begin{equation}
    \frac{\partial \rho Y_k}{\partial t} +  \frac{\partial\rho u_i Y_k}{\partial x_i}=\rho\left( \frac{\partial Y_k}{\partial t} +   u_\alpha\frac{\partial Y_k}{\partial x_i}\right) + Y_k\left( \frac{\partial \rho }{\partial t} +  \frac{\partial \rho u_i}{\partial x_i} \right).
\end{equation}
It is clear that this equation is the sum of the non-conservative form (a system characteristic) and the continuity equation. Therefore, any numerical error between the continuity equation solved by LBM and the one hidden into the conservative form leads to an inconsistency.

To summarize, provided the equations to be solved using FD/FV are only weakly coupled with the rest of the system,  the resulting hybrid LBM solver has been shown to provide reasonable results for a wide number of cases.




\paragraph{Restoring the conservativity of hybrid LBM}

Equations (\ref{eq:energy_noncons},\ref{eq:species_noncons}) are equivalent to the initial total energy and species equations (\ref{eq:energy},\ref{eq:species}), but the discrete formulation is not. This has two disadvantages: 
\begin{itemize}
    \item Global energy conservation is not numerically enforced (while the LBM scheme is numerically mass and momentum preserving).
    \item Rankine-Hugoniot relationships are not satisfied across discontinuities.
\end{itemize}

\begin{figure}[htbp]
    \centering
    \includegraphics[width=.8\textwidth]{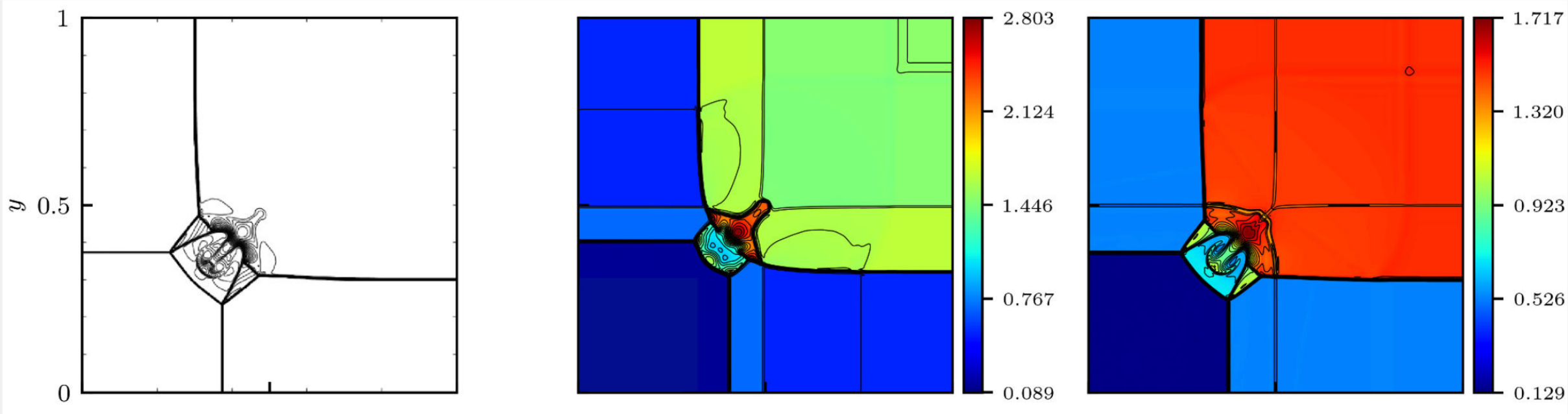}
    \caption{Two-dimensional problem of Lax \& Liu~\cite{lax_solution_1998}. From left to right: reference solution~\cite{lax_solution_1998}, solution obtained entropy equation (Muscl-Hancock), solution obtained with corresponding total energy equation scheme \cite{wissocq2022Etot}. Shown are the density fields for configuration 3 at time $t=0.3$.}
    \label{fig_Etot_Riemann}
\end{figure}

The latter is clearly visible in Fig.~\ref{fig_Etot_Riemann}, which presents a reference 2-D Riemann problem and the solution as obtained with hybrid LBM using Muscl-Hancock scheme to solve the entropy equation.

Wissocq et al. \cite{wissocq2022Etot} recently presented a method to construct a linearly equivalent total energy scheme from any FD/FV scheme that is linearly stable for hybrid LBM (e.g., for the entropy equation). 

\subsection{Compressible continuity and momentum balance equations}\label{subsec:compressibleLBM}
All strategies presented above for the resolution of the additional energy and species equations coupled to a LBM solver require modifications to the LBM core. The major options are presented hereafter.

\subsubsection{Lattices with higher-order quadratures}
\begin{figure}
	\centering
		 \includegraphics[width=5cm]{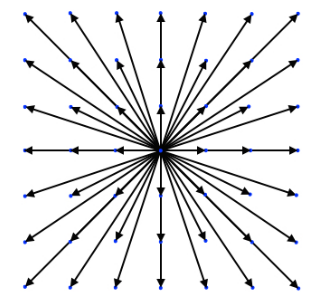}
    \caption{Illustration of the D2Q49 lattice.}
\label{Fig:d2q49_lattice}
\end{figure}

\begin{figure}
	\centering
		 \includegraphics[width=5cm]{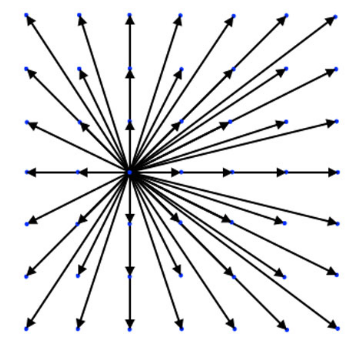}
    \caption{Illustration of the D2Q49 lattice with a shift of $U_x=\delta x/\delta t$.}
\label{Fig:d2q49_shifted_lattice}
\end{figure}

\begin{figure}
	\centering
		 \includegraphics[width=8cm]{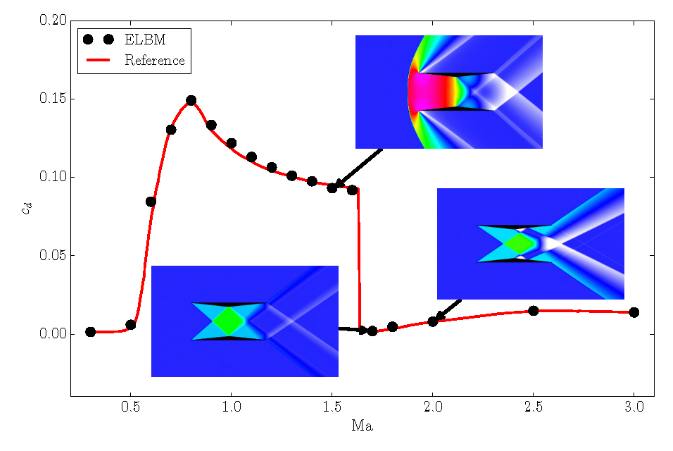}
    \caption{Drag coefficient $c_d$ as a function of the free stream Mach number for the Busemann biplane simulations. Inset: snapshots of the pressure distribution around the biplane for three different Mach numbers: Ma $= 1.5$, top; Ma $= 1.7$, bottom left; Ma $= 2.0$, bottom right. Figure reproduced from \cite{frapolli_theory_2020}. }
\label{Fig:busemann_biplane}
\end{figure}

\paragraph{Standard approach with polynomial equilibria}
As discussed in the isothermal models sections the Maxwell-Boltzmann phase-space continuous equilibrium can be matched at the discrete level via a number of methods following the same general principle, matching the different moments of the continuous equilibrium with the discretized version. As such the classical moment-matching method routinely used for Eulerian discrete velocity Boltzmann models and truncated Hermite expansion approach both fall into that category. In the case of the former one, once the number of constraints on moments of the equilibrium and degrees of freedom in the form of number of discrete velocities have been set, construction of the discrete equilibria boils down to solving the following linear system:
\begin{equation}\label{eq:polyEqSys}
    \bm{M f^{\rm eq}} = \Pi^{\mathrm{MB}},
\end{equation}
where $\Pi^{\bm MB}$ is a vector of size $1 \times Q$, $Q$ being the number of constraints on moments, with moments of the Maxwell-Boltzmann continuous distribution corresponding to the targeted constraints with:
\begin{equation}
    \Pi^{\mathrm{MB}}_{n} = \int v_{x}^{p} v_{y}^{q} v_{z}^{r} f^{\mathrm{MB}} d\bm{v},
\end{equation}
with $n=p+q+r$. The quantity $\bm{f^{\rm eq}}$ is the vector of sizes $1 \times Q$ containing discrete equilibria and $\bm{M}$ the transformation matrix from discrete equilibria to moments. For instance, in 1-D, for a solver targeting the Navier-Stokes-Fourier dynamics a minimum of five discrete velocities are needed as one must correctly recover moments of order zero to four. This approach to construct a discrete solver, while being quite flexible has a number of shortcomings, namely: The matrix $\bm{M}$ is not necessarily invertible for any choice of moments and discrete velocities as illustrated by the introduction of so-called \emph{anti-symmetric} discrete velocities in some higher-order discrete velocity Boltzmann models, see for instance \cite{gan_discrete_2018}; while the number of velocities is set by the constraints the sizes and size-ratios of these velocities have no \emph{a priori} closures and are usually tuned via trial and error. A possible closure for the size of the discrete velocities would be to use Hermite polynomials roots of the corresponding order. The only issue with that choice is that above order three Hermite roots do not guarantee space-filling lattices and therefore on-lattice propagation.\\
Nevertheless, a large number of publications using larger discrete velocity sets are documented in the literature:
\begin{itemize}
    \item A group of these publications do not rely on Lagrangian approaches to discretize physical space and time and use Eulerian approaches such as finite differences or finite volumes to discretize the coupled system of hyperbolic equations of the discrete velocity Boltzmann model. In doing so the sizes of discrete velocities can be freely tuned to stabilize the simulation for a specific test-case.
    \item Another group of publications use Lagrangian approach to discretize physical space and time and overcome the issue of non-space-filling lattices by supplementing the collision-propagation step with an interpolation step to bring back post-streaming discrete velocities on lattice. These approaches are sometimes referred to as \emph{semi-Lagrangian}, see for instance \cite{bardow_multispeed_2008,wilde_semi-lagrangian_2020}.
    \item Another category of publications, relying on the classical on-lattice method, proposes to stabilize multi-speed lattice Boltzmann solvers for compressible flows through different collision models, such as multiple-relaxation time or regularized, see for instance \cite{coreixas_recursive_2017, jacob_new_2018}.
\end{itemize}
All of the previously-listed models have had limited success in modeling generic high-speed compressible flows with large temperature variations in the domain. A number of alternatives have been proposed since then to considerably widen the stability domain of multi-speed lattice Boltzmann solvers. They will be discussed next.
\paragraph{Extension of stability domain: Entropic equilibria}
The entropic construction of the discrete equilibrium state introduced for isothermal models, can be reformulated in a more general form as a minimization problem subject to $M$ constraints:
\begin{equation}\label{eq:constrained_minimization}
    \delta H + \delta(\sum_{m=0}^{M-1} \lambda_m \Pi_m) = 0.
\end{equation}
The formal solution of this constrained minimization leads to a function of the following form:
\begin{equation}
    f_i^{\rm eq} = \rho w_i \exp{\left[\sum_{m=0}^{M} \lambda_m \left(\sum_{j=0}^{Q-1} \frac{\partial \Pi_m}{\partial f_j} \right)\right]}.
\end{equation}
Note that other form of the minimizer without the weights $w_i$ have also been proposed and used in the literature~\cite{ottinger_formulation_2020}, most notably for entropic Grad moments methods~\cite{ottinger_formulation_2020,levermore_moment_1996}.
For instance, a model imposing only constraints on collisional invariants, i.e.
\begin{subequations}
\begin{align}
    \sum_{i=0}^{Q-1} f_i^{\rm eq} &= \rho,\\
    \sum_{i=0}^{Q-1} c_{i\alpha} f_i^{\rm eq} &= \rho u_\alpha,\\
    \sum_{i=0}^{Q-1} c_{i\alpha}^2 f_i^{\rm eq} &= \rho \left(u_\alpha + D r T \right),
\end{align}
\end{subequations}
would lead to the following discrete equilibrium~\cite{frapolli_theory_2020}:
\begin{equation}
    f_i^{\rm eq} = \rho w_i \exp{\left[\lambda_0 + \lambda_\alpha c_{i\alpha} + \lambda_2 \bm{c}_i^2 \right]}.
\end{equation}
It is interesting to note that while, for the most part, entropic equilibria construction has been done by enforcing constraints on collisional invariants, one may reduce higher-order moments error by adding corresponding constraint in Eq.~\eqref{eq:constrained_minimization}. This is sometimes referred to as \emph{guiding} the equilibrium and corresponding discrete equilibria are referred to as \emph{guided equilibria}~\cite{prasianakis_lattice_2007,latt_efficient_2020}. In the context of the lattice Boltzmann method, this extension of constraints was discussed for the first time in \cite{karlin_equilibria_1998} through the concept of auxiliary and target equilibria. There, auxiliary equilibria were constructed by enforcing constraints on collisional invariants and target equilibria, a combination of auxiliary equilibria and additional degrees of freedom, by enforcing constraints on higher order moments.\\
Once the form of the equilibrium distribution function has been determined, its construction consists of finding the expression of the different Lagrangian multiplicators. This is done by introducing back the discrete equilibrium into the set of constraints which would lead to a system of $M$ equations with $M$ unknowns, i.e. the Lagrange multipliers, to be determined. While an analytical expression was derived for the isothermal case with $D+1$ constraints, for larger systems no such solutions exist. In the absence of a closed form solution one can use numerical methods such as Newton iterations to find the Lagrange multipliers at every grid-point and every time-step~\cite{frapolli_entropic_2017}.\\
As shown in previous sections, one systematic approach to choose an optimal set of discrete velocities is to rely on the Gauss-Hermite quadrature and roots of Hermite polynomials. However, apart from the third-order quadrature leading to the DdQ3$^d$ lattices, all other higher order quadratures result in off-lattice propagation of some of the discrete distribution functions. In \cite{karlin_factorization_2010}, starting from a set of discrete velocities the authors proposed an approach to find a reference temperature and corresponding weights. This is achieved through the \emph{closure relation} and \emph{matching} conditions. For a set of discrete velocities $\mathcal{V}$ with $Q$ vectors $c_i$, the $Q^{\rm th}$ power of $c_i$ can be written as a linear combination of lower order odd-powers from $Q-2$ to $1$, i.e.
\begin{equation}
    c_i^Q = a_{Q-2}c_i^{Q-2} + a_{Q-4}c_i^{Q-4} + \dots + a_1 c_i.
\end{equation}
For instance, in the case of the D1Q3 lattice one has $c_i^3 = c_i$. This essentially means that the moment of order $Q$ is not an independent moment and can not be set at one's will. The only possibility is to set the linear in $u$ term of the $Q^{\rm th}$ order to its Maxwell-Boltzmann counter-part and in doing so determine the reference temperature, which is referred to as the \emph{matching} condition. Consider for instance the D1Q3 lattice again. The third-order moment is going to be $u_x$ while the Maxwell-Boltzmann distribution leads to $u_x^3+3T_0u_x$. To match the linear term one must have $3T_0 = 1$. Note that not any choice of lattice admits a reference temperature. For example the velocity set $\mathcal{V}=\{-2,-1,0,+1,+2\}$ will lead to a closure relation of the form $c_i^5=5c_i^3-4c_i$ and a matching condition, $15 T_0^2 - 15T_0 + 4 = 0$, which does not admit any solutions. This explains why the shortest admissible five-velocity lattice is $\mathcal{V}=\{-3,-1,0,+1,+3\}$ with $T_0=1\pm\sqrt{2/5}$. Once the reference temperature is determined, the weights are readily found by matching the moments of the discrete equilibrium at $\rho=1$ and $u_x=0$ to their Maxwell-Boltzmann counter-parts. Considering the condition of positivity of the weights one also find the range of temperature that can be covered by the chosen system of discrete velocities. The closure relations and reference temperatures of a number of 1-D lattice are summarized in Table~\ref{table:minimal_lattices}.\\
\begin{table*}
\begin{tabular}{|l|l|l|l|}
\hline
$Q$ & $V$ & Closure & $T_0$                                                     \\
\hline
$3$ &  $\{0,\pm 1\}$ & $c_i^3=c_{i}$& $1/3$                               \\
\hline
$5$& $\{0,\pm 1,\pm 3$\} &  $c_i^5=10 c_i^3-9c_i$& $1\pm\sqrt{2/5}$  \\
\hline
$7$& $\{0,\pm 1,\pm 2, \pm 3\}$ & $c_i^7=14 c_i^5-49 c_i^3+36 c_i$ & $0.697953$ \\
\hline
$9$& $\{0,\pm 1,\pm 2, \pm 3, \pm 5\}$ & $c_i^9=39 c_i^7-399 c_i^5+1261 c_i^3-900 c_i$ & $0.756081$, $2.175382$\\
\hline
$11$& $\{0,\pm 1,\pm 2, \pm 3, \pm 4 \pm 5\}$ & $c_i^{11}=55 c_i^9-1023 c_i^7+7645 c_i^5-21076 c_i^3+14400 c_i$ & $ 1.062794$\\
\hline
\end{tabular}
\caption{One-dimensional Maxwell lattices with odd number of integer-valued velocities, $Q=3,5,7,9,11$.
Second column: Lattice vectors;
Third column: Closure relation, defining the reference temperature $T_0$ through the matching condition (fourth column).}
\label{table:minimal_lattices}
\end{table*}
One successful example of such lattices is the D2Q49 shown in Fig.~\ref{Fig:d2q49_lattice}.
The closure relation for the 1-D set is
\begin{equation}
    c_i^7 = 14 c_i^5 - 49 c_i^3 + 36 c_i.
\end{equation}
The 1-D weights read:
\begin{subequations}
    \begin{align}
        w_{0} &= \frac{36 - 49T+ 42T^2 - 15T^3}{36},\\
        w_{\pm1} &= \frac{T(12-13T+5T^2)}{16},\\
        w_{\pm2} &= \frac{T(-3+10T - 5 T^2)}{40},\\
        w_{\pm3} &= \frac{T(4-15T+15T^2)}{720},
    \end{align}
\end{subequations}
which lead to $T_{\rm min}=1-\sqrt{2/5}$ and $T_{\rm max}=1+\sqrt{2/5}$. Note that the range of accessible temperatures can be further extended by changing the ratio of the largest and shortest discrete velocities, here $\pm3$ and $\pm1$. In \cite{frapolli_entropic_2017} the author also proposed pruning strategies to reduce the number of discrete velocities in 2-D and 3-D, leading to the D3Q39 lattice, which reduces the discrete velocities by one order of magnitude compared to the tensor product of the D1Q7, i.e. D3Q343.
\paragraph{Adaptive reference frame models}
As observed for both isothermal and compressible models, errors in higher-order moments scale with the deviations of local temperature and velocity from the lattice reference temperature and velocity. For all symmetric lattices considered up to that point the lattice reference velocity is $U=0$. In \cite{frapolli_lattice_2016} the authors proposed to challenge the idea of a reference frame at rest by introducing a non-zero shift $U$. It was noted that the discrete entropy functional is uniquely defined by the weights $w_i$. The weights of a lattice with $Q$ discrete velocities, as shown in the previous section, are determined by matching the first $Q$ moments of the Maxwell-Boltzmann equilibrium distribution function at temperature $T$ and $u_x=0$:
\begin{equation}
    \sum_{i=0}^{Q-1} \phi(c_i) w_i(0,rT) = \int \phi(v)f^{\rm MB}(0,rT) dv.
\end{equation}
It was shown through the Galilean-invariance of the moments of the Maxwell-Boltzmann distribution function and the binomial theorem that the weights are also Galilean-invariant and therefore untouched by the change of reference frame. The immediate consequences of that observation are: (a) construction of 3-D lattices via tensorial product of the 1-D lattice remains as before, (b) assuming $U=k\delta x/\delta t$ with $k\in\mathbb{Z}$ the propagation remains on-lattice and (c) the discrete entropy functional is Galilean invariant and therefore equilibrium populations are form invariant under the shift of reference frame. This point along with the effect of the shift on operation range has also been discussed for standard isothermal lattices~\cite{hosseini_extensive_2019}. The process of changing the reference frame and resulting discrete lattice is illustrated in Fig.~\ref{Fig:d2q49_shifted_lattice} through the D2Q49 lattice.
The use of the shifted lattice along with the entropic equilibrium has been successfully used to model a wide variety of high Mach number flows as illustrated in Fig.~\ref{Fig:busemann_biplane}.
The idea of shifted reference frames was later generalized to local adaptive reference velocity and temperature
through the particles on demand (PonD) method~\cite{dorschner_particles_2018}. In this approach the collision streaming operation is performed on a reference frame corresponding to the local velocity and temperature. This allows to minimize higher-order moments deviations of the discrete equilibrium from the Maxwell-Boltzmann distribution function and in doing so allows for arbitrarily large variations in speed and temperature. The particles on demand method has been used to model high Mach number cases in recent years~\cite{kallikounis_particles_2022}. It is also currently used in combination with a Lee-Tarver reaction model to simulate detonation at high Mach numbers, see \cite{sawant_detonation_2022}.
\begin{figure}
    \centering
    \includegraphics[width=7cm,keepaspectratio]{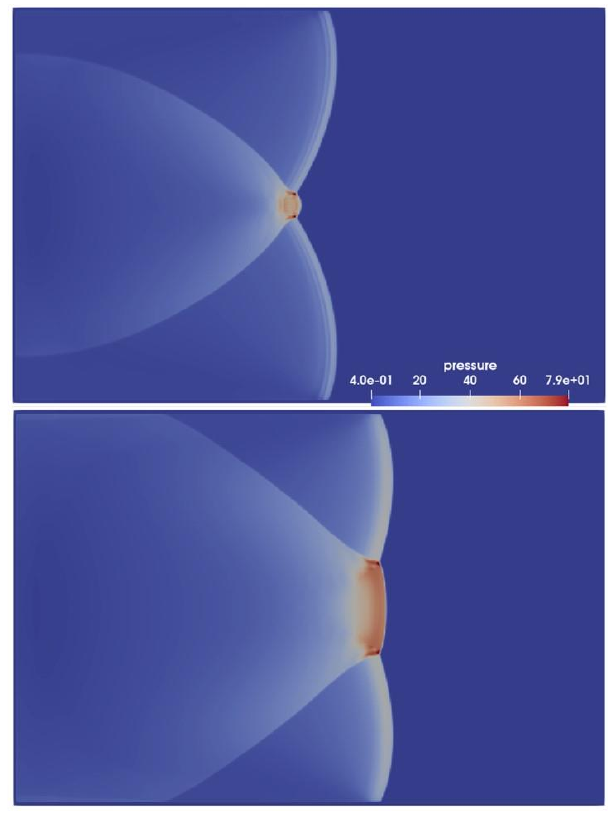}
    \caption{Mach reflection and regular reflection created from the interaction of incident detonation waves for adiabatic exponent (top) $\gamma=1.4$, or (bottom) $\gamma=5/3$. Images reproduced from \cite{sawant_detonation_2022}. }
    \label{Fig:PECS_pond_detonation}
\end{figure}
\subsubsection{Standard lattice density-based solvers}
\begin{figure}[htbp]
  \centering
    \includegraphics[width=0.4\textwidth]{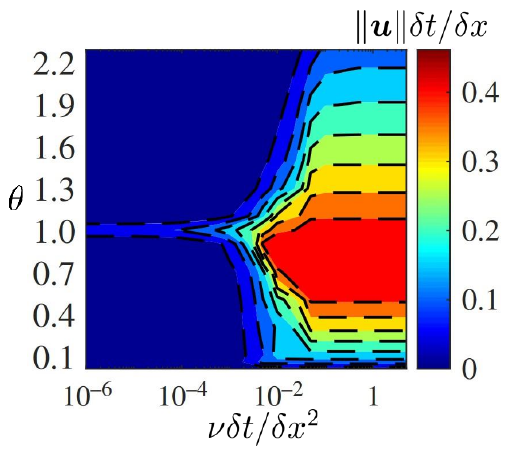}   
  \caption{Linear stability domain of lattice Boltzmann at different non-dimensional temperatures and viscosities as obtained from von Neumann analysis. Reproduced from ~\cite{hosseini_compressibility_2020}.\label{fig:thermal_stability}}
\end{figure}

\paragraph{Coupling to temperature field}
As discussed in the first section, the original lattice Boltzmann method was targeting the incompressible limit as the asymptotic behavior of the incompressible Navier-Stokes equations. To that end the temperature appearing in the equilibrium distribution function was that of a reference state guaranteeing validity of the low Mach assumption. The compressible Navier-Stokes equations can be recovered by replacing the reference temperature with the local fluid temperature obtained from the second distribution function or the FD/FV solver used for energy balance; Considering for instance Eq.~\ref{eq:prod_form_xi} it changes into:
\begin{equation}
    \zeta_{\alpha\alpha}=c_s^2\theta+u_{\alpha}^2,
\end{equation}
where $\theta = \bar{r}T/\bar{r}_0 T_0$. Introducing this term allows for a correct recovery of Euler level pressure while setting the relaxation time to:
\begin{equation}
    \bar{\tau} = \frac{\nu}{c_s^2\theta} + \frac{\delta t}{2},
\end{equation}
allows for correct recovery of the Navier-Stokes level viscous stress coefficient. With the temperature now an independent space- and time-varying parameter, it will inevitably deviate from the reference temperature, i.e. $\theta=1$ which is the optimal operation temperature of the third-order quadrature-based lattice Boltzmann model. Deviation from the reference temperature comes with a number of difficulties. The first one is the reduced domain of stability, illustrated best by the linear stability domain shown in Fig.~\ref{fig:thermal_stability}.
The second is that to properly recover the full Navier-Stokes viscous stress tensor a number of additional considerations have to be taken into account. These are discussed in the next paragraph.
\paragraph{Galilean-invariance of third-order moments and corrections}
As discussed for the isothermal lattice Boltzmann method in previous sections, a simple CE analysis shows that at the NS level, moments of orders two and three of the EDF must be correctly recovered. Diagonal components of the third-order moments tensor, i.e. moments of the form $\Pi_{\alpha\alpha\alpha}$, can not be correctly recovered due to the $c_{i\alpha}^3=c_{i\alpha}$ bias of the third-order quadrature-based lattice. While the continuous Maxwell-Boltzmann equilibrium distribution leads to:
\begin{equation}
    \Pi_{\alpha\alpha\alpha}^{\rm MB} = \rho u_{\alpha}^3 + 3\rho c_s^2 u_{\alpha}\theta,
\end{equation}
any of the discrete equilibrium distribution functions discussed here recovers:
\begin{equation}
    \Pi_{\alpha\alpha\alpha}^{\rm eq} = 3\rho c_s^2, 
\end{equation}
which for $\theta=1$ introduces a cubic-in-velocity error and for $\theta\neq1$ a linear one. As such the issue of Galilean-variance of the third-order moments becomes quite critical in the case of compressible flows where $\theta\neq{\rm const}$. To account for this error, corrections in the form of source terms in the kinetic equation are introduced:
\begin{equation}
    \partial_t f_i + c_{i\alpha}\frac{\partial f_i}{\partial x_\alpha} = \frac{1}{\tau}\left(f_i^{\rm eq} - f_i \right) + \Psi_i.
\end{equation}
The form of the source term, $\Psi_i$, can be derived through the order-two-in-$\epsilon$ (NS level) momentum balance equation:
\begin{equation}
  \frac{\partial^{(2)}\rho u_\alpha}{\partial t} + \frac{\partial}{\partial x_\beta} \tau\left(\frac{\partial^{(1)} \Pi^{\rm eq}_{\alpha\beta}}{\partial t} +\frac{\partial \Pi^{\rm eq}_{\alpha\beta\gamma}}{\partial x_\gamma}\right) \\ + \frac{\partial}{\partial x_\beta} \tau\left(\sum_{i=0}^{Q-1} c_{i\alpha}c_{i\beta}\Psi^{(1)}_i\right) = 0,
\end{equation}
leading to:
\begin{equation}\label{eq:correction_1}
  \Psi_i^{(1)} = \frac{w_i}{2c_s^4}\frac{\partial_\alpha}{\partial x_\alpha} \mathcal{H}_{\alpha\alpha}(\bm{c}_i)\delta\Pi^{\rm eq}_{\alpha\alpha\alpha},
\end{equation}
where
\begin{equation}
    \delta\Pi^{\rm eq}_{\alpha\alpha\alpha} = \rho u_\alpha \left[ u_\alpha^2 + 3 c_s^2\left(\theta-1\right)\right].
\end{equation}
For the stress tensor to be correctly recovered at this scale one must have:
\begin{equation}\label{eq:correction_1p}
  \Psi_i = \frac{w_i}{2c_s^4}\partial_\alpha\mathcal{H}_{i,\beta\gamma}\delta\Pi^{\rm eq}_{\alpha\beta\gamma}.
\end{equation}
Note that to get this expression the correction term was assumed to involve first-order derivatives via the expansion $\Psi_i=\Psi_i^{(1)}$.\\
A different form of the correction term can be obtained with a different expansion, i.e. $\Psi_i'=\Psi_i^{(2)}$. Such an expansion would lead to the following NS-level equation:
\begin{equation}
  \frac{\partial^{(2)} \rho u_\alpha}{\partial t} + \frac{\partial}{\partial x_\beta} \tau\left(\frac{\partial^{(1)} \Pi^{\rm eq}_{\alpha\beta}}{\partial t} +\frac{\partial \Pi^{\rm eq}_{\alpha\beta\gamma}}{\partial x_\gamma}\right) \\ - \sum_{i=0}^{Q-1} c_{i\alpha}{\Psi'_i}^{(2)} = 0,
\end{equation}
and a correction term of the form:
\begin{equation}\label{eq:correction_2}
    {\Psi^{'}_i} = \frac{w_i}{c_s^2}c_{i\alpha}  \frac{\partial}{\partial x_\alpha}\left(\frac{\mu}{p} \frac{\partial \delta \Pi^{\rm eq}_{\alpha\alpha\alpha}}{\partial x_\alpha}\right).
\end{equation}
The above-listed corrections were derived for the discrete kinetic equations. The classical lattice Boltzmann approach to space/time discretization would lead to the following redefined discrete distribution function:  
\begin{equation}
    \bar{f}_i = f_i - \frac{\delta t}{2}\Omega_i - \frac{\delta t}{2}\Psi_i,
\end{equation}
which in turn would lead to the following final algebraic system:
\begin{equation}
\bar{f}_i\left(\bm{x}+\bm{c}_i\delta t, t+\delta t\right) - \bar{f}_i\left(\bm{x}, t\right) = \frac{\delta t}{\bar{\tau}}\left( f^{\rm eq}_i\left(\bm{x}, t\right) - \bar{f}_i\left(\bm{x}, t\right)\right) + \left(1 - \frac{\delta t}{2\bar{\tau}}\right)\Psi_i.
\end{equation}
This consistent derivation of the \emph{extended} LBM holds for any realization of the correction term, whether it is introduced simply as a Hermite-expanded term~\cite{feng_three_2015} or by using the extended equilibrium approach~\cite{saadat_extended_2021}.
\subsubsection{Pressure-based solvers}
\begin{figure}[htbp]
  \centering
\includegraphics[width=0.4\textwidth]{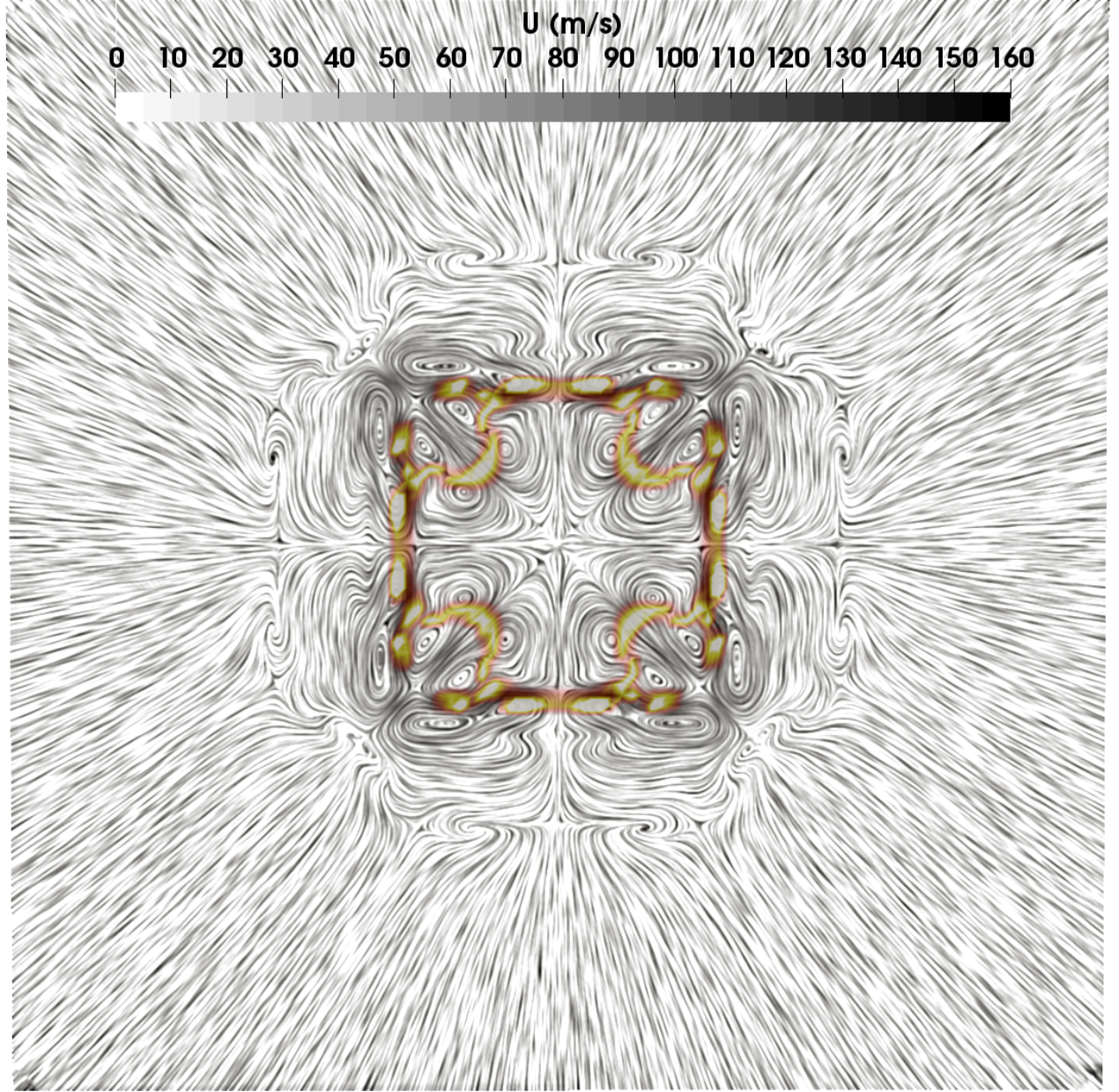} 
  \includegraphics[width=0.4\textwidth]{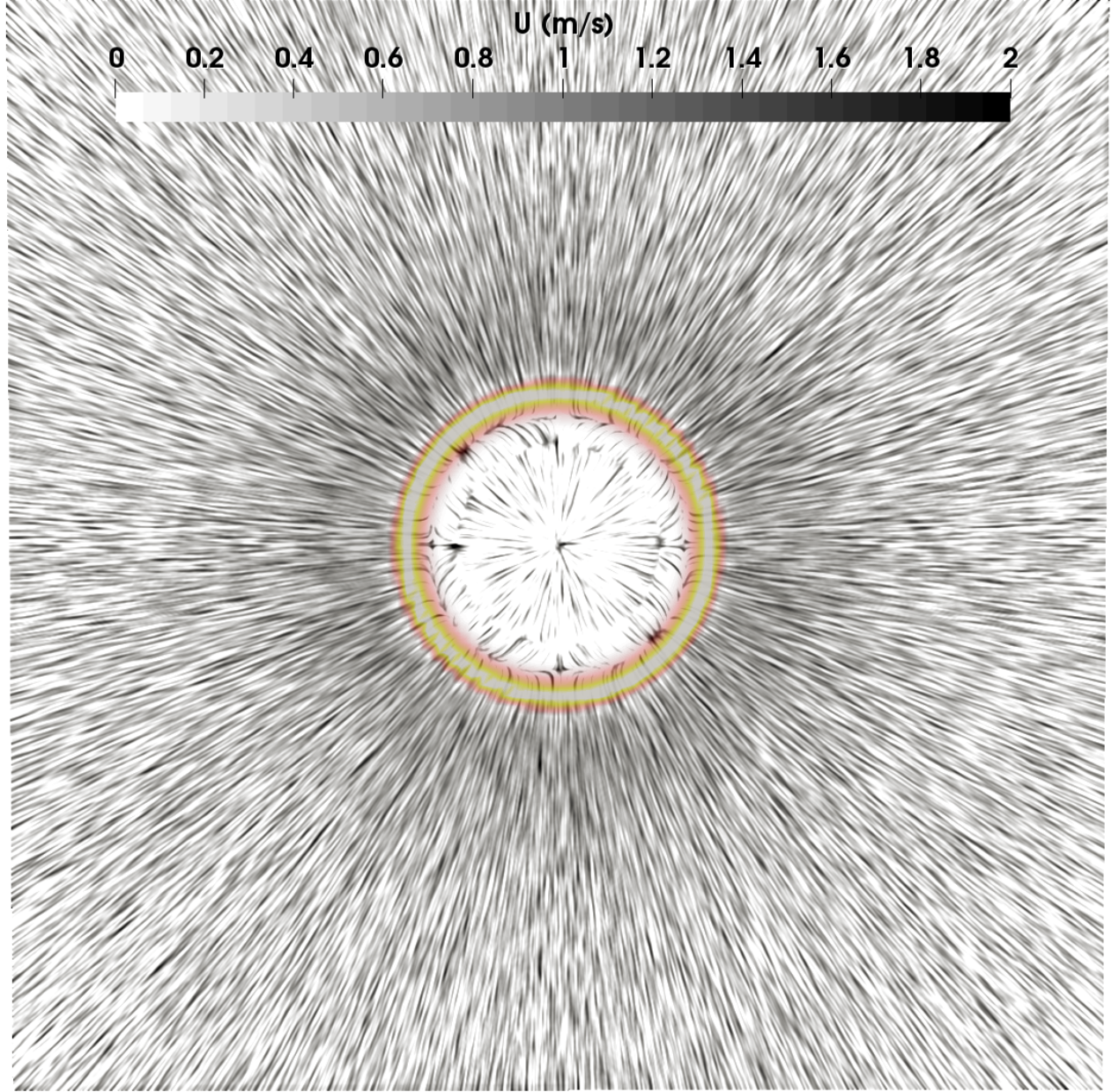} 
  \caption{Streamlines of the 2-D circular flame simulation colored by velocity magnitude~(in m/s). Left column: density-based model \cite{tayyab2020hybrid}, right column:  pressure-based model \cite{tayyab2020experimental}. Note the very different ranges of velocity magnitude from the two methods. The yellow contour is the heat release rate peak indicating the flame front. \label{fig:sf-velocomp}}
\end{figure}

While the density-based model detailed in the previous section was successfully used for a number of applications, see~\cite{feng2018combustion,tayyab2020hybrid}, it was observed that it led to spurious currents near curved flame interfaces, see Fig.~\ref{fig:sf-velocomp} for a circular flame. A detailed study of numerical properties of the model showed that this is a result of a non-physical coupling between entropic and vorticity mode, see~\cite{renard_linear_2020}.
A pressure-based formulation was then proposed \cite{farag2020pressure} to cure the problem, the detailed reason only being understood later \cite{farag2021bridge}. 

The pressure-based algorithm is presented hereafter. 
\begin{description}
    \item[Step 1] Calculation of the  $p$-based equilibrium distribution $f^{p,eq}_i$ from $(t,\boldsymbol{x})$ moments :
     \begin{equation}
        f^{p,eq}_i(t,\boldsymbol{x}) = \omega_i \Big\{ \mathcal{H}^{(0)} \rho \theta + \displaystyle\frac{\mathcal{H}^{(1)}_{i\alpha}}{c_s^2}\rho u_\alpha  + \displaystyle\frac{\mathcal{H}^{(2)}_{i\alpha\beta}}{2c_s^{4}}[\rho u_\alpha u_\beta] + \displaystyle\frac{\mathcal{H}^{(3)}_{i\alpha \beta \gamma}}{6c_s^{6}}[\rho u_\alpha u_\beta u_\gamma] \Big\}(t,\boldsymbol{x}) \,.\label{hrrp_equilibrium}
    \end{equation}
    \item[Step 2] Off-equilibrium population reconstruction  $\overline{f}^{neq}_i(t,\boldsymbol{x})$ from moments $\left[ \rho, \rho u_\alpha, \Pi_{\alpha\beta}^{neq} \right](t,\boldsymbol{x})$, using a collision model, e.g. projected \cite{latt_lattice_2006} or recursive \cite{coreixas_recursive_2017,jacob_new_2018} regularization.
    \item[Step 3] Collision and streaming,
    \begin{align}
        &f^{p,col}_i(t,\boldsymbol{x}) = f^{p,eq}_i(t,\boldsymbol{x}) + \left( 1 - \displaystyle\frac{\delta t}{\overline{\tau}} \right)  \overline{f}^{neq}_i(t,\boldsymbol{x})\,,\label{hrrp_collision}\\
       &\overline{f}^{p}_i(t+\delta t,\boldsymbol{x}) = f^{p,col}_i(t,\boldsymbol{x}-\boldsymbol{c_i}\delta t)\,.\label{hrrp_streaming}
    \end{align}
       \item[Step 4] macroscopic reconstruction 
    \begin{align}
       & \rho(t+\delta t,\boldsymbol{x}) = \displaystyle\sum_i  \overline{f}^p_i(t+\delta t,\boldsymbol{x}) + \rho(t,\boldsymbol{x})[1-\theta(t,\boldsymbol{x})].\label{hrrp_mass_scheme_corr}\\
       & \rho u_\alpha(t+\delta t,\boldsymbol{x}) = \displaystyle\sum_i  c_{i\alpha}\overline{f}^p_i(t+\delta t,\boldsymbol{x}) \,,\label{hrrp_momentum_scheme}\\
       & \Pi_{\alpha\beta}^{\overline{f}^{neq}}(t+\delta t,\boldsymbol{x}) = \displaystyle\sum_i  c_{i\alpha} c_{i\beta}\left[\overline{f}^p_i- f^{p,eq}_i\right](t+\delta t,\boldsymbol{x}) \,,\label{hrrp_stresstensor_scheme}
    \end{align}
    \item Update of the energy variable (hybrid or DDF method) \cite{feng2019hybridJCP,farag2020pressure,renard2021improved,song2020numerics}. From this additional step, $\theta(t+\delta t,\boldsymbol{x})$ is now updated.
\end{description}

Note the differences compared to the density-based model presented in the previous section: 
\begin{itemize}
    \item The zeroth moment of $f_i$ is now $p=\rho\theta$ instead of $\rho$.
    \item The macroscopic reconstruction of Step 4 now includes a correction $\rho(t+\delta t,\boldsymbol{x}) = \displaystyle\sum_i  \overline{f}^p_i(t+\delta t,\boldsymbol{x}) + \rho(t,\boldsymbol{x})[1-\theta(t,\boldsymbol{x})]$ accounting for dilatation. 
\end{itemize}
This second point was presented by the authors as a predictor-corrector procedure, close to early artificial compressibility methods \cite{chorin67,chorin68}. It is important to note, however, that despite being pressure-based, the algorithm is mass conserving globally, as density-based methods.

\paragraph{Link between pressure-based and density based formulations}
Since many mesh transition algorithms, boundary conditions, etc. were initially developed for density-based algorithms \cite{astoul2021lattice}, there is an interest in establishing a rigorous link between pressure-based and density-based algorithm. 

This can be obtained noting that the correction $\rho(t,\boldsymbol{x})[1-\theta(t,\boldsymbol{x})]$ in the macroscopic reconstruction can be equivalently embedded directly in the $f_0$ term corresponding to the stationary discrete velocity by introducing the density-based function:
\begin{equation}
    \overline{f}^{\rho}_i(t+\delta t,\boldsymbol{x}) = \overline{f}^{p}_i(t+\delta t,\boldsymbol{x}) +
    \delta_{0i}[1-\theta(t,\boldsymbol{x})],
\label{eq_pbased_rhobased}
\end{equation}
where $\delta_{0i}$ is the Kronecker symbol. By projecting $\delta_{0i}$ on a Hermite polynomial basis, it was further shown \cite{farag2021bridge} that this change is equivalent to adding to the classical $\overline{f}^{\rho}_i$ a fourth order contribution, leading to an equilibrium function of the generic form
\begin{align}
    f^{eq}_i = \omega_i \Big\{ \mathcal{H}^{(0)} \rho+ \displaystyle\frac{\mathcal{H}^{(1)}_{i\alpha}}{c_s^2}\rho u_\alpha  + \displaystyle\frac{\mathcal{H}^{(2)}_{i\alpha\beta}}{2c_s^{4}}\left[\rho u_\alpha u_\beta + \delta_{\alpha\beta}\rho c_s^2(\theta-1)\right] + \displaystyle\frac{\mathcal{H}^{(3)}_{i\alpha \beta \gamma}}{6c_s^{6}}\Big[\rho u_\alpha u_\beta u_\gamma \nonumber\\ - \kappa\rho c_s^2\left(u_\alpha \delta_{\beta\gamma} + u_\beta \delta_{\gamma\alpha} + u_\gamma \delta_{\alpha\beta}\right)\Big] - \frac{\mathcal{A}_i+\mathcal{B}_i+\mathcal{C}_i}{12c_s^4} \rho[\theta-1](1-\zeta) \Big\} \,,\label{general_feq2}
\end{align}
 with additional information projected onto fourth order polynomials $\mathcal{A}_i$, $\mathcal{B}_i$ and $\mathcal{C}_i$. In the model, $\kappa$ and $\zeta$ are free parameters. For instance, $(\kappa,\zeta)=(1,1-\theta)$ corresponds to the density-based model of \cite{renard2021improved}, while $(\kappa,\zeta)=(0,0)$ yields the pressure-based model of \eqref{hrrp_equilibrium}.

\begin{figure}[htbp]
    \centering
    \begin{minipage}{.54\textwidth}
\includegraphics[width=0.49\textwidth]{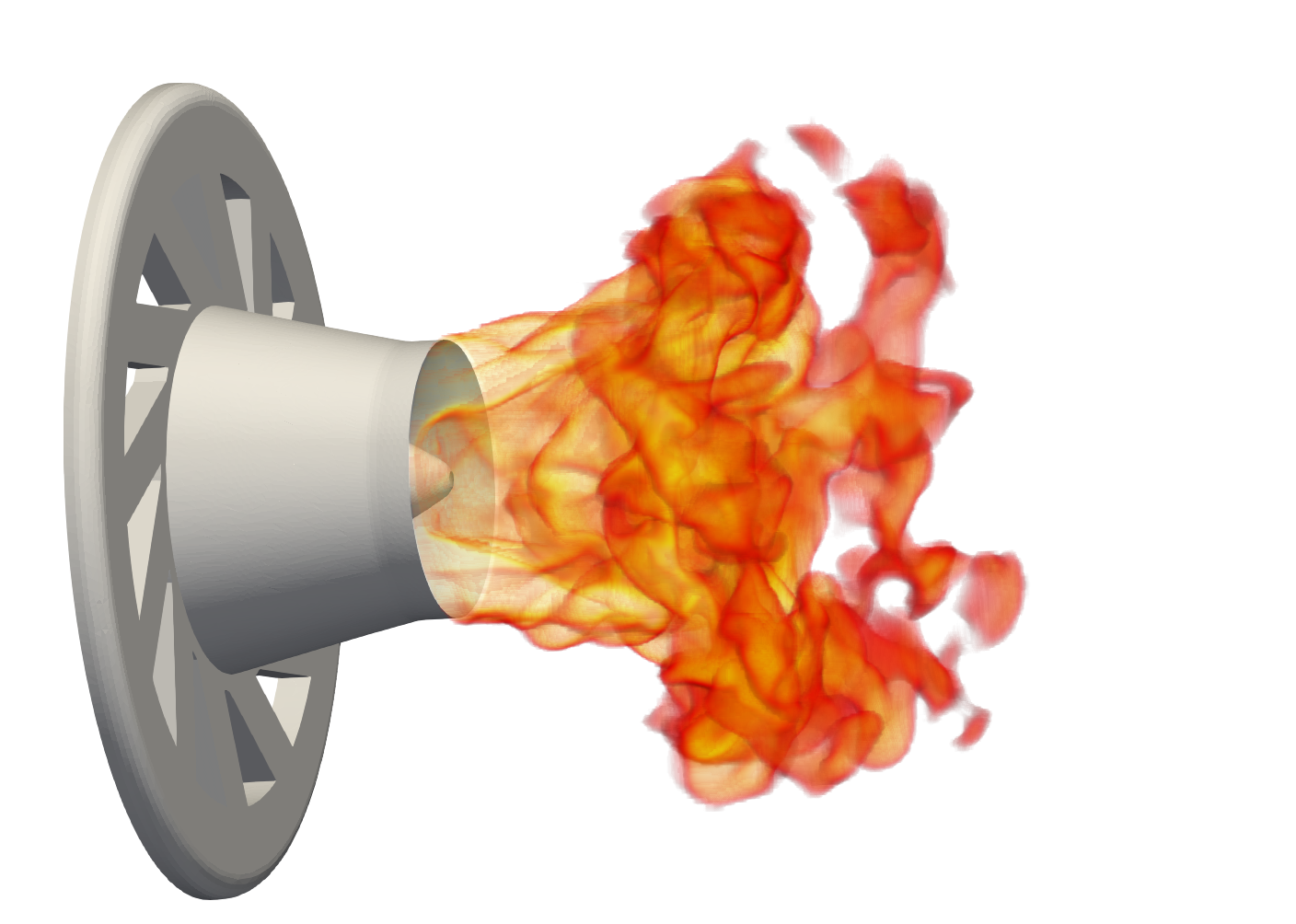}
{\includegraphics[width=0.49\textwidth]{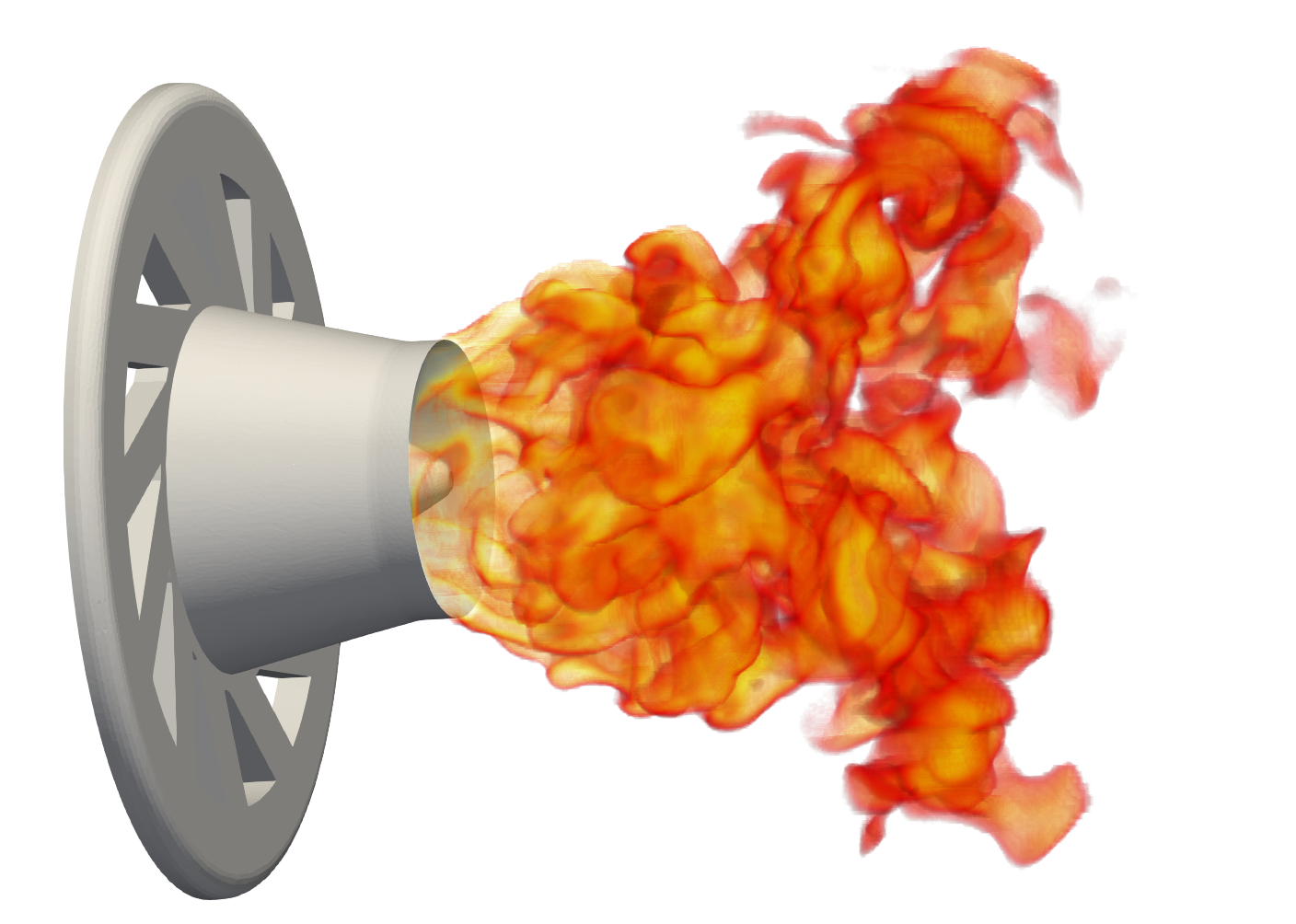}} \\
 {\includegraphics[width=0.49\textwidth]{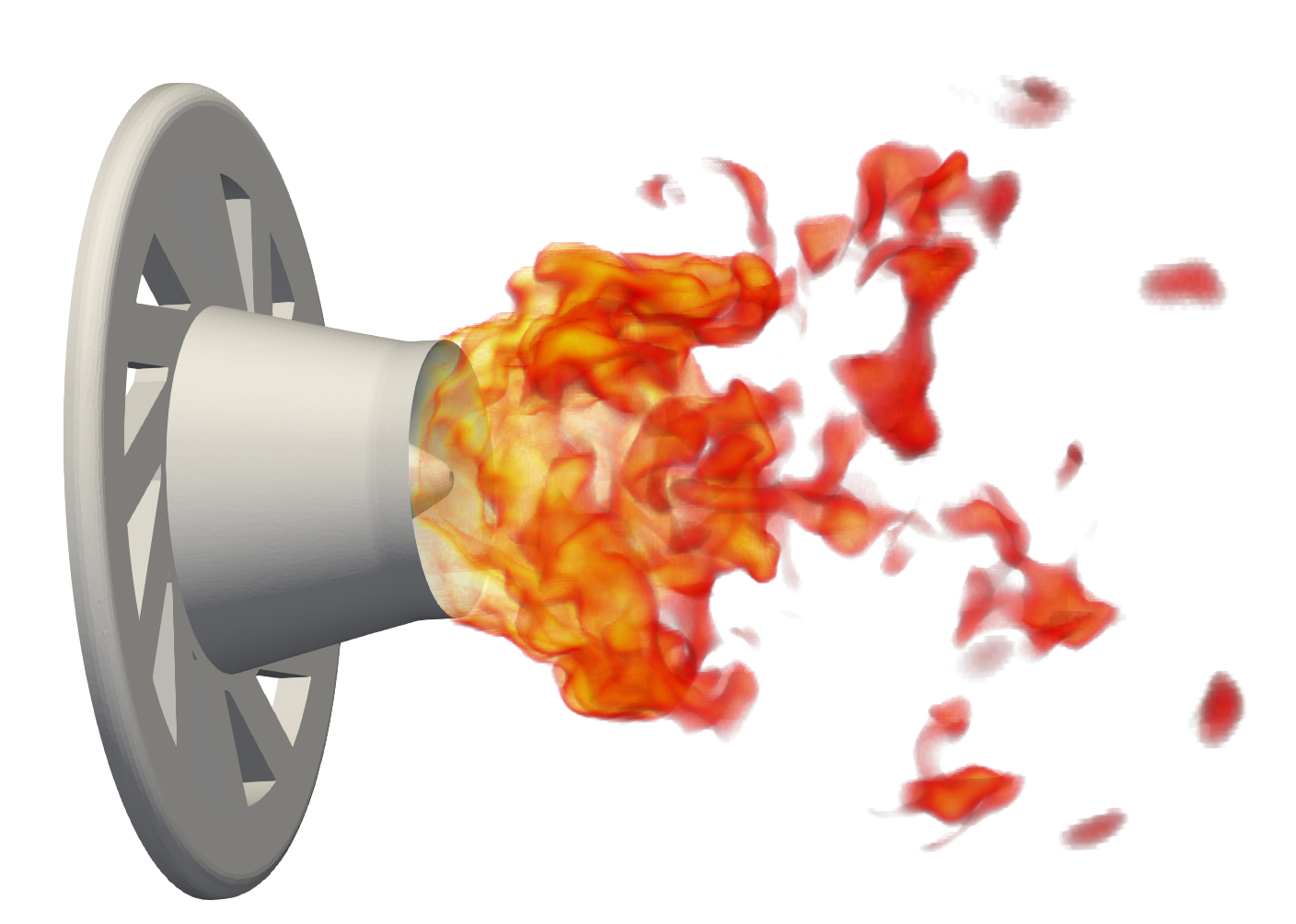}}
{\includegraphics[width=0.49\textwidth]{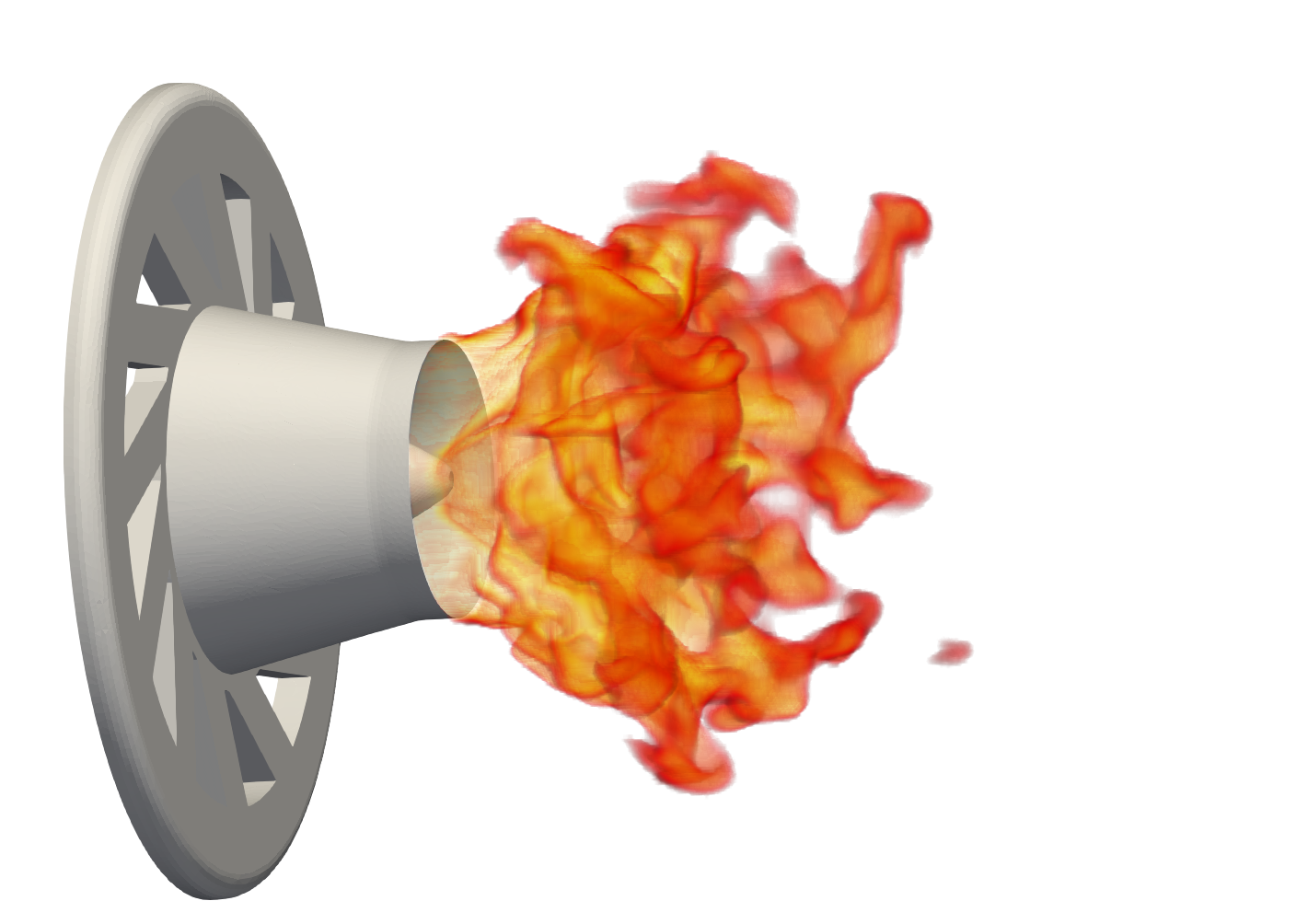}}        
    \end{minipage}
\begin{minipage}{.45\textwidth}
    \includegraphics[width=\linewidth]{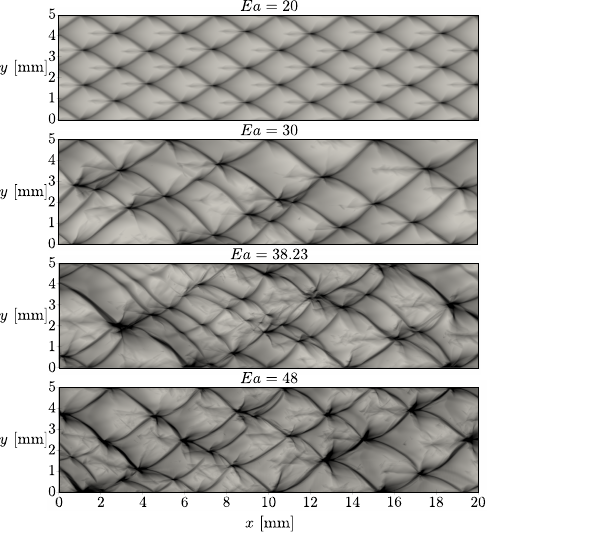}
\end{minipage}
    \caption{Illustration of successful applications of the generic HRR model \eqref{general_feq2}: (left) a cycle of a thermo-acoustic instability in the PRECCINSTA burner \cite{zhao2023preccinsta}, and (right) 2-D detonation cellular structure with varying activation energy \cite{wissocq2023deto}.}
    \label{fig:HRRgeneric}
\end{figure}

Successful applications of the resulting generic model include the modelling of
\begin{itemize}
    \item Hele-Shaw cell \cite{tayyab2020experimental};
    \item Turbulent premixed combustion burners; \cite{tayyab2021volvo,zhao2023preccinsta}
    \item Turbulent lifted \ce{H2}-air jet flame \cite{taileb2022cabra};
    \item Thermo-acoustic instabilities \cite{karthik2022lattice,zhao2023preccinsta};
    \item Cellular detonation structure \cite{wissocq2023deto};
\end{itemize}
with the last two points being illustrated in Fig.~\ref{fig:HRRgeneric}.

\subsubsection{Low Mach thermo-compressible pressure-based solver}
\begin{figure*}
    \centering
    \includegraphics[width=13cm,keepaspectratio]{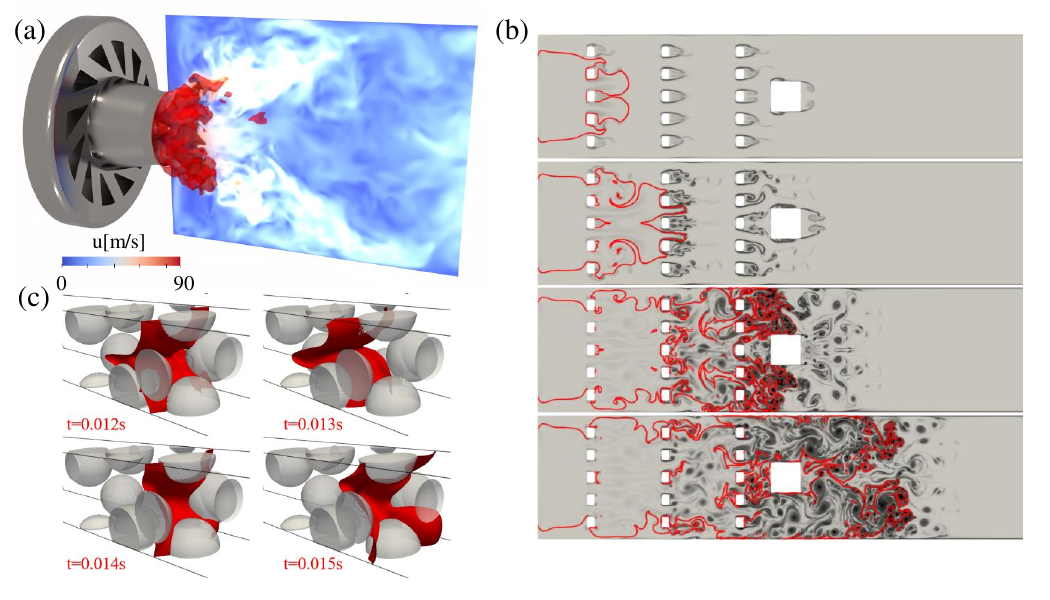}
    \caption{Illustration of some of the recent applications of the low Mach model of \cite{hosseini_hybrid_2019}. (a) Simulation of PRECCINSTA swirl injector at equivalence ratio $0.83$~\cite{hosseini_low_2022}, the red surface illustrating the flame structure. (b) Simulation of deflagrating flame in a chamber with obstacles~\cite{hosseini_low_2022}. (c) Simulation of flame propagation in randomly-generated porous media~\cite{hosseini_towards_2023}.}
    \label{Fig:PECS_LMNA_applications}
\end{figure*}

In 2019, Hosseini et al. proposed a low Mach lattice Boltzmann scheme for simulations of combustion, and more generally of dilatable flows~\cite{hosseini_hybrid_2019}. A low Mach reduction of the fully compressible models of the previous section was also proposed in \cite{wang2022lmna}. The scheme is categorized as low Mach in the sense that it follows the philosophy of Majda's zero-Mach model~\cite{majda_derivation_1985} where after a Helmholtz decomposition of the velocity field, the divergence-free part is obtained via Poisson's equation and the curl-free part from the species and energy fluxes. At the difference of Majda's zero-Mach model, here the \emph{divergence-free} component solver allows for a certain level of compressibility, i.e. spurious acoustic waves. To recover this modified set of macroscopic equations the model makes use of the following modified kinetic system~\cite{hosseini_hybrid_2019}:
\begin{equation}\label{eq:LMNA_time_evolution}
    \frac{\partial g'_i}{\partial t} + c_{i\alpha}\frac{\partial g'_i}{\partial x_\alpha} = \frac{1}{\tau}\left({g_i^{\rm eq}}' - g'_i\right) + \Xi_i,
\end{equation}
where
\begin{equation}\label{eq:LMNA_modified_distribution}
    g'_i = w_i p_h + c_s^2 \left(f_i - w_i \rho \right),
\end{equation}
and the source term $\Xi_i$ is defined as:\begin{equation}\label{eq:LMNA_source_term}
    \Xi_i =  c_s^2\left(\frac{f_i^{\rm eq}}{\rho}-w_i \right)\left(c_{i\alpha}-u_\alpha\right)\frac{\partial \rho}{\partial x_\alpha} \\ + w_i c_s^2 \rho \frac{\partial u_\alpha}{\partial x_\alpha}.
\end{equation}
In this model, hydrodynamic pressure $p_h$ and velocity $\bm{u}$ are coupled via the distribution function via:
\begin{subequations}
	\begin{align}
    \sum_{i=0}^{Q-1} g'_i &= p_h,\\
	\sum_{i=0}^{Q-1} c_{i\alpha} g'_i &= \rho c_s^2 u_\alpha,
	\end{align}
\end{subequations}
while density $\rho$ is now a variable computed locally through the ideal equation of state:
\begin{equation}
    \rho = \frac{p_{th}}{\bar{r}T}.
\end{equation}
The source of divergence appearing in Eq.~\eqref{eq:LMNA_source_term} is computed via the continuity equation combined with the energy and species balance equations:
\begin{equation}
    \frac{\partial u_\alpha}{\partial x_\alpha} = - \frac{1}{p_{th}}\frac{d p_{th}}{dt} + \frac{1}{T}\left(\frac{\partial T}{\partial t} + u_\alpha\frac{\partial T}{\partial x_\alpha}\right)  + \sum_{k=1}^{N_{sp}}\frac{\bar{W}}{W_k}\frac{1}{T}\left(\frac{\partial Y_k}{\partial t} + u_\alpha\frac{\partial Y_k}{\partial x_\alpha}\right),
\end{equation}
where summation of $\alpha$ is assumed. A multi-scale analysis of this kinetic model shows that the following balance equation is effectively applied to the hydrodynamic pressure~\cite{hosseini_development_2020}:
\begin{equation}
    \frac{1}{\rho c_s^2}\partial_t p_h + \frac{\partial u_\alpha}{\partial x_\alpha} = - \frac{1}{p_{th}}\frac{d p_{th}}{dt} + \frac{1}{T}\left(\frac{\partial T}{\partial t} + u_\alpha\frac{\partial T}{\partial x_\alpha}\right) + \sum_{k=1}^{N_{sp}}\frac{\bar{W}}{W_k}\frac{1}{T}\left(\frac{\partial Y_k}{\partial t} + u_\alpha\frac{\partial Y_k}{\partial x_\alpha}\right),
\end{equation}
while for momentum, as for the classical lattice Boltzmann with second-order equilibrium, the Navier-Stokes equation is recovered with a deviation of order $\propto {\|\bm{u}\|}^3 \delta t^3/\delta x^3$ in both diagonal and deviatoric components of the viscous stress tensor.\\
Note that after integration along characteristics, the discrete time evolution equations for the re-defined distribution function $\bar{g'}_i$ are obtained as:
\begin{equation}
    \bar{g'}_i(\bm{x}+\bm{c}_i \delta t, t+\delta t) - \bar{g'}_i(\bm{x}, t) =  \frac{\delta t}{\bar{\tau}}\left( \bar{g'}_i^{\rm eq}(\bm{x}, t) - \bar{g'}_i(\bm{x}, t)\right) + \left(1-\frac{\delta t}{2\bar{\tau}}\right)\Xi_i,
\end{equation}
while moments are computed as:
\begin{subequations}
	\begin{align}
    \sum_{i=0}^{Q-1} \bar{g'}_i + \frac{\delta t c_s^2}{2} \left( \rho \frac{\partial u_\alpha}{\partial x_\alpha} + u_\alpha \frac{\partial \rho}{\partial x_\alpha} \right)&= p_h,\\
	\sum_{i=0}^{Q-1} c_{i\alpha} \bar{g'}_i &= \rho c_s^2 u_\alpha.
	\end{align}
\end{subequations}
While only the most basic single relaxation time form of this model is introduced here, interested readers can readily get access to more advanced realizations, for instance via the Cumulants-based multiple relaxation time collision operator in~\cite{geier_cumulant_2015,hosseini_low_2022}.\\
Over the past couple of years this low Mach lattice Boltzmann solver, in combination with shock-capturing finite differences schemes like the weighted essentially non-oscillatory (WENO) have been used to model a variety of complex reacting flow configurations, including
\begin{itemize}
    \item Turbulent combustion \cite{hosseini_low-mach_2020};
    \item Combustion in swirl burner~\cite{hosseini_low_2022};
    \item Combustion in porous media~\cite{hosseini_towards_2023}.
\end{itemize}
Some of these applications are illustrated in  Fig.~\ref{Fig:PECS_LMNA_applications}. A simpler form of this model was also used in \cite{ashna_extended_2017} to model droplet combustion.
In terms of limitations, the model is -- as obvious by its name -- limited to the low Mach regime. Furthermore, as mentioned previously, the Navier-Stokes level viscous stress tensor admits deviation, for which corrections are to be published in an upcoming article. Finally, exact conservation is a topic that needs improvement here as the form of the energy and species balance equations along with the finite-difference discretizations and curved boundary treatment all lead to loss of exact conservation.
\subsection{Performance of multi-physics LBM solvers}\label{subsec:wrapup}
Let us now provide details regarding the advantages and limitations of multi-physics LBM solvers, and how they compare to classical LBM solvers targeting mainly low-Mach aerodynamic and aero-acoustic applications. 

Classical LBM solvers have found their success due to three main reasons : (i) ability to tackle complex geometries in a simple way, (ii) low dissipation owing to the streaming step, and (iii) a reduced computational cost due to the stencil compactness and octree structure. These advantages come at the cost of a heavier memory load (with more degrees of freedom), a complex treatment of boundary conditions (owing to the non body-fitted mesh), and filtering problems in  grid refinement areas where both spatial and time discretizations are halved/doubled.

Let us now consider the three aforementioned advantages and see if they apply to LBM multiphysics solvers. (i) Ability to tackle complex geometries is preserved, as the discretization remains identical. (ii) The low dissipation property is more model dependent. For multiple distribution functions, each distribution is solved using the same algorithm so that dissipation is supposed to remain identical. For hybrid approaches, however, a separate numerical scheme is used for energy and species equations, which may lead to additional dissipation \cite{farag2020pressure} on the entropy/enthalpy Kovasznay modes. 
Comparing the computational cost (iii) between LBM and Navier-Stokes solvers is a long standing issue. While a thorough comparison is still lacking, some LBM studies report reduced computational times (RCT), consistently below 5$\mu$s per time step and grid point for relevant combustion problems \cite{boivin2021tgv,tayyab2021volvo,zhao2023preccinsta,abdelsamie2021taylor}.

We list in Tab. \ref{tab_list_problems} different references tackling combustion problems using multi-physics LB solvers (regardless of the strategy). This fast-expanding list is a clear indication that multi-physics LBM solvers have now reached sufficient maturity for combustion applications.
\begin{table}[htbp]
    \centering
    \begin{tabular}{l|c}
        Canonical configuration & References  \\ \hline
        Laminar premixed and diffusion flame & \cite{chen_novel_2007,hosseini_hybrid_2019,feng2018combustion,hosseini_development_2020}\\
        Reacting Taylor-Green vortex & \cite{boivin2021tgv,hosseini_low-mach_2020}\\
        Circular expanding flames & \cite{farag2021taylor}\\
        Darrieus-Landau instabilities & \cite{tayyab2020hybrid,tayyab2020experimental} \\
        Thermo-acoustic instabilities &  \cite{karthik2022lattice,zhao2023preccinsta} \\
        Turbulent premixed burner (LES) & \cite{tayyab2021volvo,zhao2023preccinsta,hosseini_low_2022} \\
        Flame in porous media & \cite{hosseini_towards_2023}\\
        Detonations & \cite{wissocq2023deto,sawant_detonation_2022}
    \end{tabular}
    \caption{A list of canonical combustion problems treated using lattice Boltzmann methods in the recent literature}
    \label{tab_list_problems}
\end{table}

\section{Conclusion and discussion}
While extension of the lattice Boltzmann method to the simulation of combustion happened slower than other areas of application such as incompressible and multi-phase flows, the progress documented in recent years has laid the ground for widespread use and application of lattice Boltzmann to combustion. The complex simulations reported in the literature, see for instance~\cite{hosseini_low_2022,hosseini_towards_2023,zhao2023preccinsta,wissocq2023deto}, show that the numerical approach has reached a level of maturity allowing it to be applied to realistic configurations. In this contribution we focused on the development of lattice Boltzmann-based approaches to model combustion and discussed some of the most pressing challenges and solutions proposed in the literature.\\
The evolution of the literature shows that one of the major challenges preventing progress in that area was the development of stable and efficient solutions to extend the lattice Boltzmann solvers to compressible regimes. Stability has been one of the most restricting issues in that area. While use of higher order lattices is one straightforward approach to move up to compressible flows, it has been observed that apart from the additional memory and computation costs stemming from the larger number of discrete velocities, higher-order quadrature are subject to more restrictive stability conditions, especially on the temperature. This has led the community to opt for approaches relying on low-order quadratures, i.e. third-order, which had shortcomings regarding the Navier-Stokes-level viscous stress tensor due to insufficient degrees of freedom in the model. Introduction of correction terms for the viscous stress tensor along with more robust collision operators has now paved the way for simulations involving large temperature variations.\\
Other issues specific to the simulation of combustion in the context of the lattice Boltzmann method are tied to additional balance equations for species and energy. A number of different strategies have been devised for these additional fields. Some rely on developing kinetic models and, therefore, lattice Boltzmann solvers for multi-species fluids, while others prefer either lattice-Boltzmann-based passive scalar solvers or classical FV/FD solvers for the balance equations, leading to a hybrid formulation.\\
While state-of-the-art lattice Boltzmann solvers are now routinely used for complex combustion configurations involving complex geometries and turbulent flows, a number of technical challenges still persist:
\begin{itemize}
    \item One of the remaining challenges is to get exactly conservative curved boundary conditions for the lattice Boltzmann solver. While the bare half-way bounce-back method resulting in a stair-case approximation to the geometry~\cite{ladd_numerical_1994,ladd_lattice-boltzmann_2001} ensures mass conservation, all curved treatment presented in the literature, see for instance~\cite{filippova_lattice-boltzmann_1997,bouzidi_momentum_2001,mei_accurate_1999}, result in loss of conservativity of the boundary condition. For a more detailed discussed on conservation issues for curved boundary treatment interested readers are referred to~\cite{lallemand_lattice_2003,sanjeevi_choice_2018,xu_theoretical_2022}. A number of routes can be taken to overcome this issue such as the use of immersed boundaries which would come at the cost of diffuse interfaces, or the use of volumetric/fluxed based boundary treatments, see for instance \cite{chen_realization_1998,rohde_volumetric_2002}.
    \item Development and implementation of conservative and efficient dynamic grid-refinement strategies is also another topic to be further developed in the future. Although grid-refinement in the context of the lattice Boltzmann method has been developed and used since the early 2000's, see for instance~\cite{dupuis_theory_2003,rohde_generic_2006,chen_grid_2006,dorschner_grid_2016,astoul_analysis_2020}, mass-conservation, spurious currents at refinement interfaces and dynamic refinement are still topics of discussion in the literature.
\end{itemize}
At the end of this detailed review regarding past achievements obtained for combustion thanks to LB simulations, it is now time to look briefly to the future. This work is obviously not the first review concerning lattice Boltzmann simulations, and previous studies have sometimes included long-term perspectives, most prominently in the work by Succi~\cite{succi_lattice_2015} -- recently updated in~\cite{succi_lattice_2023}. It is obviously necessary to evaluate again recent evolutions in the light of those predictions. One must keep in mind that applications are in the focus of the present review, as stated in the title, so that it is does not appear meaningful to include exceedingly exotic concepts here.

In the same manner, the future of high-fidelity combustion simulations has been discussed in previous reviews, for instance~\cite{trisjono_systematic_2015,vervisch_best_2016,giusti_turbulent_2019}. Here also, reflecting on the corresponding statements will be useful. Of course, the aspects already discussed at length in the core of the present article will not be repeated here in the interest of space. Since the focus has been placed on important methodological aspects of LB in this review, a bird's view seems more appropriate to finish. As such, the main points emerging for the foreseeable future would read as follows.
\begin{itemize}
\item Coupling of different model approaches: even after having reviewed the many advantages of LB in this article, it remains clear that alternative model formulations may be very attractive from a numerical or an algorithmic point of view for specific applications. The coupling of LB to finite differences or finite volumes has been extensively described in the section concerning hybrid models. In a similar manner, coupling LB to a variety of other approaches is feasible and attractive for corresponding problems, also when involving reacting flows. Integrating the Immersed Boundary Method (IBM) within a LB framework is particularly relevant for reactive particles~\cite{jiang_hydrodynamic_2022}. To take into account even more complex physical interactions within particulate flows, a coupling of LB with the Discrete Element Method (DEM) is the next logical step~\cite{maier_coupling_2021}. For configurations involving clearly separated fluid regions, adding a Volume of Fluid (VOF) approach -- or related methods like Volume of Pixel -- appears as a promising solution~\cite{kashani_lattice_2022}.
\item Multi-physics and multiscale applications, adaptivity: the growing performance of existing computational platforms coming with the ever-increasing complexity of the target applications, problems involving a variety of physical and chemical processes -- mostly coupled in a strongly non-linear manner -- and taking place over a broad diversity of scales in time and space progressively become the rule. While concentrating here on combustion applications, it must still be recognized that LB has now virtually been used successfully for almost any kind of flows (and even beyond fluid mechanics). The next step -- certainly soon to be reached -- will for example involve accurate LB simulations of turbulent multiphase reacting flows including realistic chemistry, thermodynamics, and transport models for species and heat, up to radiative heat transfer.\\
Such multi-physics configurations typically involve a broad range of relevant scales in time and space~\cite{van_den_akker_lattice_2018}, from micrometers to centimeters for living beings~\cite{falcucci_extreme_2021}, or even ``from inside protons to the outer Universe'', citing the optimistic statement of~\cite{succi_lattice_2023}. In that case, an optimal combination of different numerical approaches will become necessary to allow for a numerical solution of the full configuration, for instance by coupling LB to Molecular Dynamics simulations~\cite{guzman_espresso_2019,yang_multiscale_2020,xie_multiscale_2023}. Multiscale issues can be partly mitigated by using adaptivity -- in particular in space for LB (local grid refinement/coarsening), a solution that has already been discussed in this review~\cite{bellotti_multidimensional_2022,zhao2023preccinsta}.
\item Integration of Machine Learning: the incredibly fast development of Machine Learning (ML) approaches will obviously not leave LB untouched, since it is increasingly supporting all approaches used for flow simulations. Regarding more specifically LB for combustion applications -- usually taking place in the turbulent regime -- two lines of research will certainly be followed in the near future. One will concentrate on the thermoreactive part of the problem, for instance using Deep Neural Networks to describe kinetics, potentially leading to very large savings in computing time and memory~\cite{chi_efficient_2022}. The other line will concentrate on ML at subgrid scale (SGS) when using LB for Large-Eddy Simulations (LES) of turbulent reacting flows~\cite{hosseini_low_2022,zhao2023preccinsta}. In that case, either the behaviour of the pure turbulent flow will be described by a SGS model based on ML~\cite{guo2020improved} -- an approach now well established in conventional CFD~\cite{park_toward_2021}; or ML could be used to represent directly turbulent/combustion coupling at subgrid scale, an even more challenging but potentially more rewarding solution, for which much remains to be done~\cite{shin_data-driven_2021}.
\item New computational architectures: Though this might not be immediately clear for young scientists, the performance of a numerical approach is not constrained only by considerations from applied mathematics (stability, convergence, dissipation, dispersion), but is also directly impacted by the details of the computational architecture on which large-scale simulations are carried out. In that sense, the comparison of different methods in terms of computational performance completely depends on the employed system. While method 1 might be order-of-magnitude faster than method 2 on a conventional, single-core system, the comparison might be completely reversed on a large, fine-grain parallel computer, or when computing on Graphical Processing Units (GPU). In that sense, an essential advantage of LB (in addition to the error-free streaming step and linearity of the basic operator) is its locality. In the standard LB formulation, only first-neighbour communications are requested, making it perfectly suited for the currently employed computer architectures; this is one essential explanation to understand the growing success of LB for many applications. Nobody knows which computer architecture will dominate in 20 years. In his review from 2015, Succi~\cite{succi_lattice_2015} already mentioned the suitability of LB for Quantum Computing (QC). Indeed, porting high-order CFD methods involving unstructured grids on QC systems sounds like a nightmare, and LB could here again profit from its apparent simplicity and locality. Lattice Boltzmann on Quantum Computers is a subject of current research~\cite{itani_analysis_2022,ljubomir_quantum_2022}. Still, QC systems being currently barely available for researchers, the impact of Quantum Computing on future LB simulations cannot be reliably estimated. The same applies to even more exotic architectures like biological computers, that have not even entered a preliminary test-phase. 
\end{itemize}
\section*{Acknowledgements}
S.A.H. and I.K. would like to acknowledge the financial support of European Research Council (ERC) through the Advanced Grant no. 834763-PonD and computational resources provided by the Swiss National Supercomputing Center CSCS under grant no. s1212. P.B. acknowledges financial support from the French National Research agency (grants ANR-20-CE05-0009 \& ANR-21-CE05-0028), as well as computational resources provided by Centre de Calcul Intensif d’Aix-Marseille and GENCI, France (Grant A0132B11951). D.T. acknowledges financial support from the Deutsche Forschungsgemeinschaft (DFG, German Research Foundation) in TRR 287 (Project-ID No. 422037413).
\bibliographystyle{elsarticle-num} 
\bibliography{final_refs}

\end{document}